\documentclass[prb,aps,english,onecolumn,notitlepage]{revtex4-2}
\usepackage{epsfig}
\usepackage{amsmath}
\usepackage{mathtools}
\usepackage{xcolor}

\DeclarePairedDelimiterX\braket[2]{\langle}{\rangle}{#1 \delimsize\vert #2}
\usepackage{sidecap}
\usepackage[figuresleft]{rotating}
\usepackage{caption}
\usepackage{float}
\usepackage{appendix}
\usepackage{subfigure}
\usepackage{graphicx}
\usepackage{hyperref}

\begin{document}
 \title{ Exciting odd frequency equal-spin-triplet correlations at metal-superconductor interfaces}
\author{Subhajit Pal}
\author{Colin Benjamin} \email{colin.nano@gmail.com}
\affiliation{School of Physical Sciences, National Institute of Science Education \& Research, HBNI, Jatni-752050, India }
\begin{abstract}
We predict the occurrence of odd frequency equal spin-triplet correlations at normal metal-superconductor junction. This result is significant because equal spin-triplet correlations are associated with the presence of dissipation-less pure spin current. Inserting a spin-flipper at the interface of a normal metal-superconductor junction, excites equal spin triplet correlations.The existence of odd frequency equal spin-triplet correlations in absence of odd frequency mixed spin-triplet correlations is the main take home message of this work. It tallies well with the measured local magnetization density of states and spin-polarized local density of states at the interface. The importance of spin-flip scattering to the obtained results is manifest when we compare our normal metal-spin flipper-superconductor junction to other hybrid junctions where either only spin mixing or both spin mixing and spin flip scattering are present.
\end{abstract}
\maketitle
\section{Introduction}
The symmetry of the Cooper pair is intrinsic to the nature of superconductivity. From Fermi-Dirac statistics, Cooper pair wave function or pairing amplitude is anti-symmetric under exchange of all quantum numbers: time (or frequency), spin, and orbital coordinates. Pairing occurs, in general, between electrons at equal times.  This type of pairing leads to even frequency, spin-singlet, and even parity (ESE) state or even frequency, spin-triplet, and odd parity (ETO) state where odd or even denotes the orbital part of the Cooper pair wave-function. $s$ and $d$ wave pairing are examples of ESE pairing, while $p$ wave pairing is an example of ETO symmetry\cite{sig}. Pairing, surprisingly, may also occur at different times too or at finite frequency, first noticed in Ref.~\cite{bere} in $^{3}He$ and then predicted to occur in disordered superconductors\cite{kirk,beli} also. This finite frequency pairing implies an odd frequency superconductor with either odd frequency, spin-singlet, and odd parity (OSO) or odd frequency, spin-triplet, and even parity (OTE) pairing. Odd-frequency superconductivity implies that the two Cooper pair electrons are odd in the relative time coordinate or frequency. An odd frequency OSO pairing state may exist in a conventional spin-singlet superconductor\cite{bala}. Recently, odd frequency superconductivity has been predicted to occur in a host of different systems\cite{asa,kom,lko,bla,triol}, in addition to driven systems\cite{ctr,tri}.

Odd-frequency superconducting pairing can also be induced in hybrid systems such as normal-superconductor junction\cite{tan,ska,pbu,ivb,goll,esc,mst}, superconductor-ferromagnet junctions\cite{fsb,afv,berr,fom,yokoy,aan,buz,mes,anw,tsk,esch,vis,jds,mgb,kalc,robi,mess,lind,sun}, as well as topological insulators-superconductor junctions\cite{black,scha,blu,cayy,fke,kuzm,dbr,ddk,flec}. In NS junctions odd frequency pairing arises because spatial parity may be broken at interface leading to transition from even s-wave to odd p-wave symmetry\cite{amb}. Further, odd frequency pairing has been seen in systems with Rashba spin-orbit coupling\cite{bokk,amb,ree}. Experimentally, odd frequency pairing enables long range superconducting correlations, seen in ferromagnet-superconductor junctions\cite{adbe,vtp}, as also Ref.~\cite{jl}. Moreover, there is a deep relationship between odd frequency correlation and topological superconductors which might host Majorana fermions(MF's)\cite{lut,clk,lf,zian}. MF's are particles which are their own antiparticle and have great attraction due to their potential applications in topological quantum computation\cite{nay,sar}. For a MF, the normal propagator ($G_{ee}^{r}$) which describes the propagation of free electrons and the anomalous propagator ($G_{eh}^{r}$) which describes dynamics of Cooper pairs are same, $G_{ee}^{r}(\omega_{m})=G_{eh}^{r}(\omega_{m})$\cite{trio}, where $\omega_{m}$ is Matsubara frequency. Further, since $G_{eh}^{r}(\omega_{m})=1/(i\omega_{m})$ for a MF, the pair amplitude ($G_{eh}^{r}$) for an isolated MF is necessarily odd in frequency\cite{lee,tam,tamu}.

This paper predicts that odd frequency equal spin-triplet pairing can be induced in a Normal metal (N)-$s$-wave Superconductor (S) junction due to interface spin-flip scattering. To date, most predictions of odd frequency pairing in NS junctions have been associated with either spin-singlet pairing or mixed spin-triplet pairing, see Refs.~\cite{LINDER,moz}. Odd frequency mixed spin-triplet correlations have been predicted to occur in varied hybrid superconducting junctions, such as a magnetic interface in an NS junction\cite{LINDER}{or a thin ferromagnetic layer at NS interface\cite{moz}}, Kondo-type impurity embedded in s-wave superconductor\cite{kuz} and randomly embedded magnetic impurities in an s-wave superconductor\cite{fla}. A recent experimental paper on a single embedded magnetic impurity in an s-wave superconductor also sees odd frequency mixed spin-triplet correlations\cite{vpe}. {In two recent works\cite{LINDER, moz}, odd frequency mixed spin-triplet correlations occur via the spin mixing process. Spin mixing and spin-flip scattering are two different processes. In the spin mixing process, an electron experiences spin-dependent phase shifts\cite{lind}, while in the spin-flip scattering process, an electron flips its spin\cite{AJP}. A thin ferromagnetic layer at the NS interface can only generate spin mixing; it can not generate any spin-flip scattering. However, a spin flipper at the NS interface causes spin-flip scattering, and it can not create any spin mixing. We will discuss this in more detail in section V. Further, in Refs.~\cite{kuz,fla} odd frequency mixed spin-triplet correlations arise due to a magnetic impurity, similar to what happens in NFS junction wherein only spin mixing occurs. However, in our work, spin-flip scattering at the NS interface induces even and odd frequency equal spin-triplet correlations.}

What differentiates this paper from the aforesaid is that we see exclusively odd frequency equal spin-triplet correlations with absence of odd frequency mixed spin triplet correlations, which are not seen in any of these works. Odd frequency \textbf{equal} spin-triplet correlations have not been predicted as yet in any other work on Normal Metal-s wave superconductor junction except this. We cannot emphasize more the importance of observing odd frequency equal spin-triplet correlations in an s-wave superconductor. This result effectively means that we have turned an s-wave superconductor into a p-wave superconductor via doping a spin flipper at the interface of an NS junction. A hallmark of a p-wave superconductor is an equal spin-triplet pairing of its Cooper pair. Examples of p-wave superconductors are $Sr_{2}RuO_{4}$, which are exotic and difficult to work with but are predicted to host Majorana fermions. However, inducing spin-triplet p wave pairing in an s-wave superconductor would imply that generating and detecting Majorana Fermions could become much easier.

We calculate the even and odd frequency spin-singlet and triplet pairing correlations in normal metal and superconducting regions using Green's function method. Locally only, even frequency, spin-singlet, and even parity (ESE), odd frequency, equal spin-triplet, and even parity correlations(OTE-equal) are finite. While non locally, both even and odd frequency spin-singlet and equal spin-triplet correlations are non zero. We determine the spin-polarized local density of states (SPLDOS) and find its relationship with odd frequency pairing. SPLDOS both decays and oscillates in the presence of spin-flip scattering, and it matches well with the odd frequency equal spin-triplet correlations. Non locally, we see that both even and odd frequency equal spin-triplet correlations are finite in the presence of spin-flip scattering.

Finally, we comment on the significance and applications of our results. Spin-triplet correlations have increasingly been speculated to play a significant role, especially where transport via a ferromagnet is concerned. Since this kind of transport isn't affected by the exchange field of the ferromagnet, this kind of correlation can penetrate over long distances, up to hundreds of nanometers, inside the ferromagnet. But, the question is how to create such equal spin-triplet states? In our work, we answer this question in a ballistic NS junction by just putting a spin flipper at the junction interface. Spin flip scattering at junction interface induces equal spin-triplet pairing in such junctions. Equal spin-triplet correlation supports dissipationless pure spin current, which is generally long-range in nature and is of great interest in superconducting spintronics\cite{mess,lind}. Dissipationless pure spin current reduces power consumption by several orders of magnitude in ultralow-power computers\cite{hol}.

The remainder of the paper is organized as follows. In section II, we first introduce our model and discuss the theoretical background to our study by writing Hamiltonian, wave functions, and boundary conditions needed to calculate Green's functions. Section III discusses the method to calculate the induced superconducting correlations and SPLDOS from retarded Green's functions. We analyze our results for pairing correlations in normal metal and superconducting regions at zero temperature and discuss the relationship between odd frequency pairing and SPLDOS in section IV. Section V outlines how and why we get uniquely odd frequency equal spin-triplet correlations with vanishing odd frequency mixed spin-triplet correlations in our chosen normal metal-spin flipper-superconductor junction. We also compare our system with both NFS and F$_{1}$F$_{2}$S junctions wherein either only spin mixing or both spin-flip and spin mixing occur. Finally, we conclude in section VI. Analytical expressions for Green's functions and the effect of finite temperature on correlations are provided in Appendix.
\section{Scattering at Normal metal-Spin flipper-Superconductor interface}
\subsection{Hamiltonian}
We consider a 1D NS junction as shown in Fig.~1, with a spin flipper at the interface ($x=0$) and solve the scattering problem\cite{BTK}. The Hamiltonian for spin flipper, from Refs.~\cite{AJP,Liu,Maru,FC,ysr} is
\begin{equation}
H_{\mbox{Spin flipper}}=-J_{0}\delta(x)\vec{s}.\vec{S},
\end{equation}
with $J_{0}$- strength of the exchange coupling, $\vec{s}$- spin of electron/hole, and $\vec{S}$- spin of spin flipper.
In our work, the spin flipper is a delta potential magnetic impurity that can be treated similarly to an Anderson impurity. In Eq.~(1), spin flipper Hamiltonian is an effective Heisenberg term that reduces the two-electron problem to a one-electron problem\cite{Maru}. Since our problem is not a time-dependent problem, we solve it using a time-independent Schr\"{o}dinger equation modified by BdG Hamiltonian. Thus spin flipper has no dynamics of its own, differentiating our model spin-flipper from a Kondo-like magnetic impurity\cite{kuz}, which has its dynamics, and it leads to the screening of impurity spins by metallic electrons below Kondo temperature\cite{avb}. One should note that Kondo-like magnetic impurity embedded in an s-wave superconductor induces odd frequency mixed spin-triplet correlations\cite{kuz}. In our setup, on the other hand,  the spin flipper generates odd frequency equal spin-triplet correlations.

An electron/hole with spin up/down incident from the metallic(N) region interacts with the spin flipper at the interface which may result in a mutual spin flip. Electron/hole can be reflected back to N region I with spin up or down. Electron-like and hole-like quasi-particles with spin up or down are transmitted into the superconducting(S) region II for energies above the gap. Spin flipper can be thought of as a point like magnetic impurity, see Ref.~\cite{AJP}. Model Hamiltonian in Bogoliubov-de Gennes (BdG) formalism of our system as shown in Fig.~1 is given as
\begin{equation}
H_{BdG}(x)=
\begin{pmatrix}
H\hat{I} & i\Delta \Theta(x)\hat{\sigma}_{y} \\
-i\Delta^{*}\Theta(x)\hat{\sigma}_{y} & -H\hat{I}
\end{pmatrix},
\label{ham}
\end{equation}
where $H=p^2/2m^\star-J_{0}\delta(x)\vec s.\vec S -E_{F}$, $\Delta$ is superconducting gap for s-wave superconductor and $\Theta$ is Heaviside step function. First term in $H$ is kinetic energy of an electron/hole with effective mass $m^{*}$, second term describes the exchange interaction $J_{0}$ between electron/hole spin ($\vec{s}$) and spin flipper's spin ($\vec{S}$), $\hat{I}$ is identity matrix, $\hat{\sigma}$ is the Pauli spin matrix and $E_F$ is Fermi energy. Strength of exchange coupling can be expressed via dimensionless parameter $J=\frac{m^{\star}J_{0}}{ k_{F}}$, see~\cite{AJP}, as the product $J_{0}\delta(x)\vec s.\vec S $ has dimensions of energy, $\vec s$ which represents spin angular momentum of electron is in units of $\hbar$ and  $\vec S$ the spin angular momentum of spin-flipper is also in units of $\hbar$, and $\delta(x)$ having dimensions of $1/Length$, therefore $J_0$- the exchange interaction has dimensions of $Energy-length/\hbar^2$.
\begin{figure}[h]
\centering{\includegraphics[width=.35\linewidth]{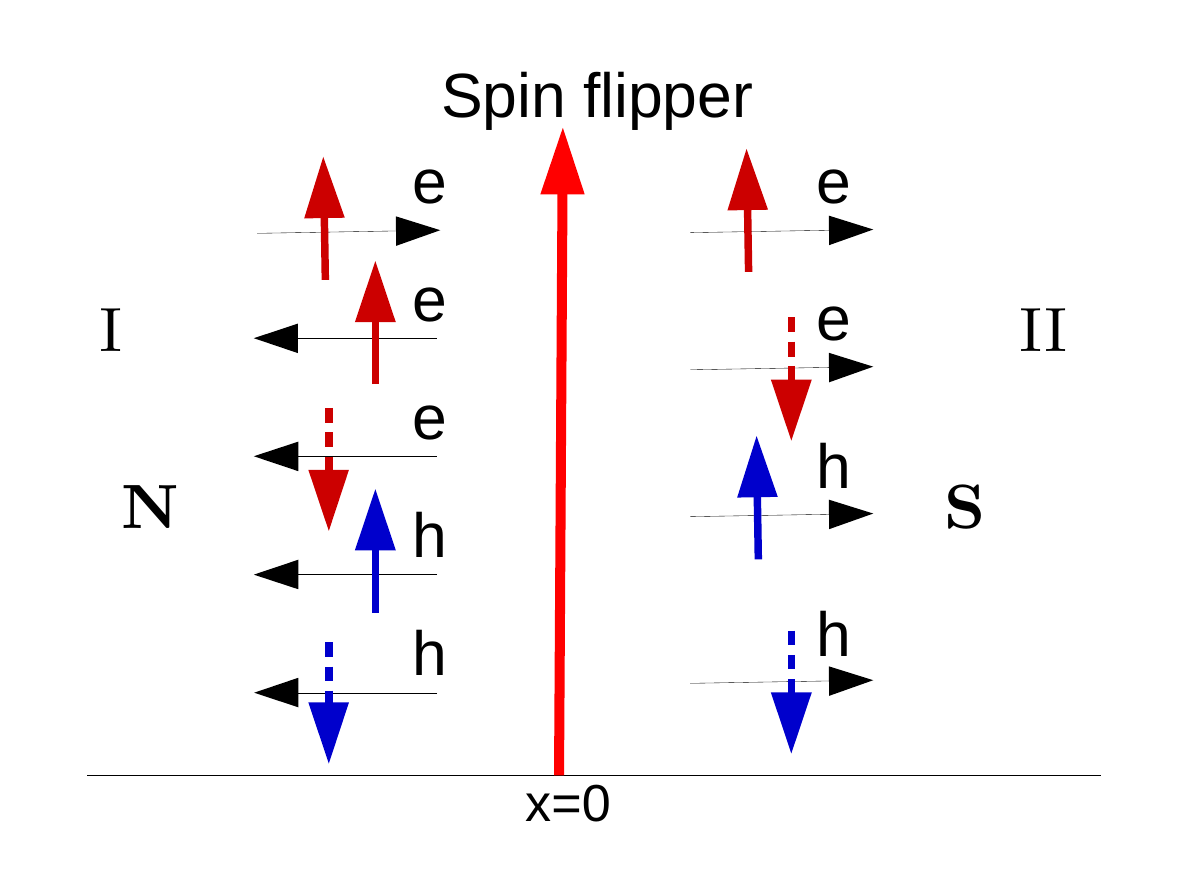}}
\caption{\small \sl NS junction with spin flipper (spin $S$ and magnetic moment $m'$) at $x=0$. Scattering of a spin up electron incident is shown. Normal reflection, Andreev reflection and quasi-particle transmission into superconductor are depicted.}
\end{figure}
\subsection{Wavefunctions}
If we diagonalize BdG Hamiltonian (Eq.~\eqref{ham}) we will get wavefunctions in different regions of our system for various types of scattering processes. Wavefunctions for different types of scattering processes are given as-
\begin{equation}
\begin{split}
\Psi_{1}(x)&=
\begin{cases}
\varphi_{1}^{N}{\rm e}^{ik_{e}x}\phi_{m'}^{S}+b_{11}\varphi_{1}^{N}{\rm e}^{-ik_{e}x}\phi_{m'}^{S}+b_{12}\varphi_{2}^{N}{\rm e}^{-ik_{e}x}\phi_{m'+1}^{S}+a_{11}\varphi_{3}^{N}{\rm e}^{i{k}_{h}x}\phi_{m'+1}^{S}+a_{12}\varphi_{4}^{N}{\rm e}^{i{k}_{h}x}\phi_{m'}^{S}\,, & x<0, \\
c_{11}\varphi_{1}^{S}{\rm e}^{ik^{S}_{e}x}\phi_{m'}^{S}+c_{12}\varphi_{2}^{S}{\rm e}^{ik^{S}_{e}x}\phi_{m'+1}^{S}+d_{11}\varphi_{3}^{S}{\rm e}^{-ik^{S}_{h}x}\phi_{m'+1}^{S}+d_{12}\varphi_{4}^{S}{\rm e}^{-ik^{S}_{h}x}\phi_{m'}^{S}
\,,& x>0.
\end{cases}
\\
\Psi_{2}(x)&=
\begin{cases}
\varphi_{2}^{N}{\rm e}^{ik_{e}x}\phi_{m'}^{S}+b_{21}\varphi_{1}^{N}{\rm e}^{-ik_{e}x}\phi_{m'-1}^{S}+b_{22}\varphi_{2}^{N}{\rm e}^{-ik_{e}x}\phi_{m'}^{S}+a_{21}\varphi_{3}^{N}{\rm e}^{i{k}_{h}x}\phi_{m'}^{S}+a_{22}\varphi_{4}^{N}{\rm e}^{i{k}_{h}x}\phi_{m'-1}^{S}\,, & x<0, \\
c_{21}\varphi_{1}^{S}{\rm e}^{ik^{S}_{e}x}\phi_{m'-1}^{S}+c_{22}\varphi_{2}^{S}{\rm e}^{ik^{S}_{e}x}\phi_{m'}^{S}+d_{21}\varphi_{3}^{S}{\rm e}^{-ik^{S}_{h}x}\phi_{m'}^{S}+d_{22}\varphi_{4}^{S}{\rm e}^{-ik^{S}_{h}x}\phi_{m'-1}^{S} \,,& x>0.
\end{cases}\\
\Psi_{3}(x)&=
\begin{cases}
\varphi_{3}^{N}{\rm e}^{-i{k}_{h}x}\phi_{m'}^{S}+a_{31}\varphi_{1}^{N}{\rm e}^{-i{k}_{e}x}\phi_{m'-1}^{S}+a_{32}\varphi_{2}^{N}{\rm e}^{-i{k}_{e}x}\phi_{m'}^{S}+b_{31}\varphi_{3}^{N}{\rm e}^{ik_{h}x}\phi_{m'}^{S}+b_{32}\varphi_{4}^{N}{\rm e}^{ik_{h}x}\phi_{m'-1}^{S}\,, & x<0, \\
c_{31}\varphi_{1}^{S}{\rm e}^{ik^{S}_{e}x}\phi_{m'-1}^{S}+c_{32}\varphi_{2}^{S}{\rm e}^{ik^{S}_{e}x}\phi_{m'}^{S}+d_{31}\varphi_{3}^{S}{\rm e}^{-ik^{S}_{h}x}\phi_{m'}^{S}+d_{32}\varphi_{4}^{S}{\rm e}^{-ik^{S}_{h}x}\phi_{m'-1}^{S}\,,& x>0.
\end{cases}\\
\Psi_{4}(x)&=
\begin{cases}
\varphi_{4}^{N}{\rm e}^{-i{k}_{h}x}\phi_{m'}^{S}+a_{41}\varphi_{1}^{N}{\rm e}^{-i{k}_{e}x}\phi_{m'}^{S}+a_{42}\varphi_{2}^{N}{\rm e}^{-i{k}_{e}x}\phi_{m'+1}^{S}+b_{41}\varphi_{3}^{N}{\rm e}^{ik_{h}x}\phi_{m'+1}^{S}+b_{42}\varphi_{4}^{N}{\rm e}^{ik_{h}x}\phi_{m'}^{S}\,, & x<0, \\
c_{41}\varphi_{1}^{S}{\rm e}^{ik^{S}_{e}x}\phi_{m'}^{S}+c_{42}\varphi_{2}^{S}{\rm e}^{ik^{S}_{e}x}\phi_{m'+1}^{S}+d_{41}\varphi_{3}^{S}{\rm e}^{-ik^{S}_{h}x}\phi_{m'+1}^{S}+d_{42}\varphi_{4}^{S}{\rm e}^{-ik^{S}_{h}x}\phi_{m'}^{S}\,,& x>0.
\end{cases}
\\
\Psi_{5}(x)&=
\begin{cases}
c_{51}\varphi_{1}^{N}{\rm e}^{-ik_{e}x}\phi_{m'}^{S}+c_{52}\varphi_{2}^{N}{\rm e}^{-ik_{e}x}\phi_{m'+1}^{S}+d_{51}\varphi_{3}^{N}{\rm e}^{i{k}_{h}x}\phi_{m'+1}^{S}+d_{52}\varphi_{4}^{N}{\rm e}^{i{k}_{h}x}\phi_{m'}^{S}\,, & x<0, \\
\varphi_{1}^{S}{\rm e}^{-ik^{S}_{e}x}\phi_{m'}^{S}+b_{51}\varphi_{1}^{S}{\rm e}^{ik^{S}_{e}x}\phi_{m'}^{S}+b_{52}\varphi_{2}^{S}{\rm e}^{ik^{S}_{e}x}\phi_{m'+1}^{S}+a_{51}\varphi_{3}^{S}{\rm e}^{-ik^{S}_{h}x}\phi_{m'+1}^{S}+a_{52}\varphi_{4}^{S}{\rm e}^{-ik^{S}_{h}x}\phi_{m'}^{S}\,,& x>0.
\end{cases}
\\
\Psi_{6}(x)&=
\begin{cases}
c_{61}\varphi_{1}^{N}{\rm e}^{-ik_{e}x}\phi_{m'-1}^{S}+c_{62}\varphi_{2}^{N}{\rm e}^{-ik_{e}x}\phi_{m'}^{S}+d_{61}\varphi_{3}^{N}{\rm e}^{i{k}_{h}x}\phi_{m'}^{S}+d_{62}\varphi_{4}^{N}{\rm e}^{i{k}_{h}x}\phi_{m'-1}^{S}\,, & x<0, \\
\varphi_{2}^{S}{\rm e}^{-ik^{S}_{e}x}\phi_{m'}^{S}+b_{61}\varphi_{1}^{S}{\rm e}^{ik^{S}_{e}x}\phi_{m'-1}^{S}+b_{62}\varphi_{2}^{S}{\rm e}^{ik^{S}_{e}x}\phi_{m'}^{S}+a_{61}\varphi_{3}^{S}{\rm e}^{-ik^{S}_{h}x}\phi_{m'}^{S}+a_{62}\varphi_{4}^{S}{\rm e}^{-ik^{S}_{h}x}\phi_{m'-1}^{S}\,,& x>0.
\end{cases}\\
\Psi_{7}(x)&=
\begin{cases}
c_{71}\varphi_{1}^{N}{\rm e}^{-i{k}_{e}x}\phi_{m'-1}^{S}+c_{72}\varphi_{2}^{N}{\rm e}^{-i{k}_{e}x}\phi_{m'}^{S}+d_{71}\varphi_{3}^{N}{\rm e}^{ik_{h}x}\phi_{m'}^{S}+d_{72}\varphi_{4}^{N}{\rm e}^{ik_{h}x}\phi_{m'-1}^{S}\,, & x<0, \\
\varphi_{3}^{S}{\rm e}^{ik^{S}_{h}x}\phi_{m'}^{S}+a_{71}\varphi_{1}^{S}{\rm e}^{ik^{S}_{e}x}\phi_{m'-1}^{S}+a_{72}\varphi_{2}^{S}{\rm e}^{ik^{S}_{e}x}\phi_{m'}^{S}+b_{71}\varphi_{3}^{S}{\rm e}^{-ik^{S}_{h}x}\phi_{m'}^{S}+b_{72}\varphi_{4}^{S}{\rm e}^{-ik^{S}_{h}x}\phi_{m'-1}^{S}\,,& x>0.
\end{cases}
\\
\Psi_{8}(x)&=
\begin{cases}
c_{81}\varphi_{1}^{N}{\rm e}^{-i{k}_{e}x}\phi_{m'}^{S}+c_{82}\varphi_{2}^{N}{\rm e}^{-i{k}_{e}x}\phi_{m'+1}^{S}+d_{81}\varphi_{3}^{N}{\rm e}^{ik_{h}x}\phi_{m'+1}^{S}+d_{82}\varphi_{4}^{N}{\rm e}^{ik_{h}x}\phi_{m'}^{S}\,, & x<0, \\
\varphi_{4}^{S}{\rm e}^{ik^{S}_{h}x}\phi_{m'}^{S}+a_{81}\varphi_{1}^{S}{\rm e}^{ik^{S}_{e}x}\phi_{m'}^{S}+a_{82}\varphi_{2}^{S}{\rm e}^{ik^{S}_{e}x}\phi_{m'+1}^{S}+b_{81}\varphi_{3}^{S}{\rm e}^{-ik^{S}_{h}x}\phi_{m'+1}^{S}+b_{82}\varphi_{4}^{S}{\rm e}^{-ik^{S}_{h}x}\phi_{m'}^{S}\,,& x>0.
\end{cases}
\end{split}
\label{wave}
\end{equation}
where $\varphi_{1}^{N}=\begin{pmatrix}
1\\
0\\
0\\
0
\end{pmatrix}$, $\varphi_{2}^{N}=\begin{pmatrix}
0\\
1\\
0\\
0
\end{pmatrix}$, $\varphi_{3}^{N}=\begin{pmatrix}
0\\
0\\
1\\
0
\end{pmatrix}$, $\varphi_{4}^{N}=\begin{pmatrix}
0\\
0\\
0\\
1
\end{pmatrix}$, $\varphi_{1}^{S}=\begin{pmatrix}
u\\
0\\
0\\
v
\end{pmatrix}$, $\varphi_{2}^{S}=\begin{pmatrix}
0\\
u\\
-v\\
0
\end{pmatrix}$, $\varphi_{3}^{S}=\begin{pmatrix}
0\\
-v\\
u\\
0
\end{pmatrix}$ and $\varphi_{4}^{S}=\begin{pmatrix}
v\\
0\\
0\\
u
\end{pmatrix}$.
$\Psi_{1}$, $\Psi_{2}$, $\Psi_{3}$ and $\Psi_{4}$ represent scattering processes when spin up electron, spin down electron, spin up hole and spin down hole are incident from N region, while $\Psi_{5}$, $\Psi_{6}$, $\Psi_{7}$ and $\Psi_{8}$ represent scattering processes when spin up electron, spin down electron, spin up hole and spin down hole are incident from S region respectively. $b_{ij}$ and $a_{ij}$ are normal reflection amplitudes and Andreev reflection amplitudes respectively, while $c_{ij}$ and $d_{ij}$ are transmission amplitudes for electron-like quasi-particles and hole-like quasi-particles respectively. $\phi_{m'}^{S}$ represents the eigenspinor for spin flipper with its $S^{z}$ operator acting as- $S_{z}\phi_{m'}^{S} = m'\phi_{m'}^{S}$, with $m'$ being spin magnetic moment of the spin flipper. $u=\sqrt{\frac{1}{2}(1+\frac{\sqrt{\omega^2-\Delta^2}}{\omega})}$ and $v=\sqrt{\frac{1}{2}(1-\frac{\sqrt{\omega^2-\Delta^2}}{\omega})}$ are BCS coherence factors. $k_{e,h}=\sqrt{\frac{2m^{*}}{\hbar^2}(E_{F}\pm\omega)}$ are wave-vectors in normal metal, while $k_{e,h}^{S}=\sqrt{\frac{2m^{*}}{\hbar^2}(E_{F}\pm\sqrt{\omega^2-\Delta^2})}$ are wave-vectors in superconductor. Conjugated processes $\tilde{\Psi_{i}}$ required to construct Green's functions in next section are obtained by diagonalizing Hamiltonian $H_{BdG}^{*}(-k)$ instead of $H_{BdG}(k)$. In case of Normal metal-Spin flipper-Superconductor junction (Fig.~1) we find that $\tilde{\varphi_{i}}^{N(S)}=\varphi_{i}^{N(S)}$ and hence $\tilde{\Psi_{i}}=\Psi_{i}$, also resulting in identical scattering amplitude, e.g., $\tilde{b}_{i1}=b_{i1}$ and so on. In our work in limit of $E_{F}>>\Delta,\omega$ we approximate $k_{e,h}\approx k_{F}(1\pm\frac{\omega}{2E_{F}})$ with $k_{F}=\sqrt{2m^{*}E_{F}/\hbar^2}$ and $k_{e,h}^{S}\approx k_{F}\pm i\kappa$ with $\kappa=\sqrt{\Delta^2-\omega^2}[k_{F}/(2E_{F})]$. Further, the superconducting coherence length\cite{ath} is given by $\xi=\hbar/(m^{*}\Delta$).
\subsubsection{Boundary condition}
Scattering amplitudes are determined by imposing boundary conditions, which at $x=0$ are \begin{eqnarray}
{}&\Psi_{i}(x<0)=\Psi_{i}(x>0),\label{boun1}\\
&\mbox { and, }\frac{d\Psi_{i}(x>0)}{dx}-\frac{d\Psi_{i}(x<0)}{dx}=-\frac{2m^{\star}J_{0}\vec s.\vec S}{\hbar^2} \Psi_{i}(x=0),
\label{boun2}
\end{eqnarray}
where $\vec s.\vec S=s_{z}S_{z}+\frac{1}{2}(s^{-}S^{+}+s^{+}S^{-})$ is exchange operator in Hamiltonian\cite{AJP}, $s^{\pm}=s_{x}\pm is_{y}$ are raising and lowering spin operator for electron and $S^{\pm}=S_{x}\pm iS_{y}$ are raising and lowering spin operator for spin flipper with $s_{z}=\frac{\hbar}{2}\begin{pmatrix}\sigma_{z} & 0 \\
0 & -\sigma_{z} \end{pmatrix}$, $s_{x}=\frac{\hbar}{2}\begin{pmatrix}  0 & \sigma_{x}\\
\sigma_{x} & 0 \end{pmatrix}$, $s_{y}=\frac{\hbar}{2}\begin{pmatrix}  0 & -i\sigma_{y}\\
i\sigma_{y} & 0 \end{pmatrix}$, $s^{+}=s_{x}+is_{y}=\frac{\hbar}{2}\begin{pmatrix}
0 & \sigma_{x}+\sigma_{y}\\
\sigma_{x}-\sigma_{y} & 0 \end{pmatrix}$ and, $s^{-}=s_{x}-is_{y}=\frac{\hbar}{2}\begin{pmatrix}
0 & \sigma_{x}-\sigma_{y}\\
\sigma_{x}+\sigma_{y} & 0 \end{pmatrix}$. $\sigma_{z}=\begin{pmatrix}1 & 0 \\
0 & -1 \end{pmatrix}$, $\sigma_{x}=\begin{pmatrix}0 & 1 \\
1 & 0 \end{pmatrix}$ and, $\sigma_{y}=\begin{pmatrix}0 & -i \\
i & 0 \end{pmatrix}$ are the usual Pauli spin matrices. Action of  exchange operator $\vec s.\vec S$, from boundary condition at $x=0$, gives for wave-function involving spin up electron spinor,
\begin{equation}
\label{EUU}
\vec s.\vec S\varphi_{1}^{N}\phi_{m'}^{S}=s_{z}S_{z}\varphi_{1}^{N}\phi_{m'}^{S}+\frac{1}{2}s^{-}S^{+}
              \varphi_{1}^{N}\phi_{m'}^{S}+\frac{1}{2}s^{+}S^{-}\varphi_{1}^{N}\phi_{m'}^{S}.
\end{equation}
Now, $s^{+}\varphi_{1}^{N}=0$, since $s^{+}$ is the spin raising operator for electron and there are no higher spin states for a spin-$1/2$ electron than up and so the 3rd term in Eq.~\ref{EUU} vanishes, while $s^{-}\varphi_{1}^{N}=\varphi_{2}^{N}$, the spin lowering operator gives the down spin state $\varphi_{2}^{N}$ of electron.  Further, for spin-up electron $s_{z}\varphi_{1}^{N}= \frac{\hbar}{2}\varphi_{1}^{N}$, and for spin flipper: $S_{z}\phi_{m'}^{S}=m'\phi_{m'}^{S}$. The spin-raising and spin-lowering operators acting on spin flipper give: $S^{+}\phi_{m'}^{S}=\sqrt{(S-m')(S+m'+1)}\phi_{m'+1}^S=F\phi_{m'+1}^S$ and $S^{-}\phi_{m'+1}^{S}=\sqrt{(S-m')(S+m'+1)}\phi_{m'}^S=F\phi_{m'}^S$.
\begin{equation}
\mbox{Thus,}\,\,
\vec s.\vec S\varphi_{1}^{N}\phi_{m'}^{S}=\frac{m'}{2}\varphi_{1}^{N}\phi_{m'}^{S}+\frac{F}{2}\varphi_{2}^{N}\phi_{m'+1}^{S}.
             \label{eu}
             \end{equation}
Similarly, action of $\vec{s}.\vec{S}$ for wave-function involving spin-down electron spinor is
\begin{equation}
\label{ed}
\vec s.\vec S \varphi_{2}^{N}\phi_{m'}^{S}=-\frac{m'}{2}\varphi_{2}^{N}\phi_{m'}^{S}+\frac{F'}{2}\varphi_{1}^{N}\phi_{m'-1}^{S}.
\end{equation}
Further, action of exchange operator for wave-function involving spin up hole is
\begin{equation}
\label{hu}
\vec s.\vec S\varphi_{3}^{N}\phi_{m'}^{S}=-\frac{m'}{2}\varphi_{3}^{N}\phi_{m'}^{S}+\frac{F'}{2}\varphi_{4}^{N}\phi_{m'-1}^{S},
\end{equation}
and finally action of exchange operator on wavefunction involving spin down hole is
\begin{equation}
\label{hd}
\vec s.\vec S\varphi_{4}^{N}\phi_{m'}^{S}=\frac{m'}{2}\varphi_{4}^{N}\phi_{m'}^{S}+\frac{F}{2}\varphi_{3}^{N}\phi_{m'+1}^{S}.
\end{equation}
In Eqs.~(\ref{eu}-\ref{hd}), $F=\sqrt{(S-m')(S+m'+1)}$ denotes spin-flip probability of spin flipper when spin up electron or spin down hole is incident, while $F'=\sqrt{(S+m')(S-m'+1)}$ denotes spin-flip probability of spin flipper when spin down electron or spin up hole is incident and $\varphi_{1}^{N}$, $\varphi_{2}^{N}$, $\varphi_{3}^{N}$, $\varphi_{4}^{N}$ are defined in Eq.~\eqref{wave}. Using above equations and solving boundary condition at $x=0$ we get 8 equations for each scattering process, see Eq.~\eqref{wave}. From each set of these $8$ equations we can calculate different scattering amplitudes: $b_{ij}$, $a_{ij}$, $c_{ij}$, $d_{ij}$. In the next section we will use these scattering amplitudes to compute retarded Green's function in each region of our system. From retarded Green's function we can calculate the induced pairing correlations, e.g., ESE, ETO, OSO, OTE, and SPLDOS in each region of junction.
\section{Green's function}
Motivation of our work is to see if via effect of spin flip scattering one can induce odd frequency spin triplet pairing in our setup. For this purpose, we follow Refs.~\cite{cayy, amb} and set up retarded Green's function $G^{r}(x,x',\omega)$ with outgoing boundary conditions in both N and S regions due to interface\cite{mcm} scattering. Retarded Green's function then is
\begin{equation}
\label{RGF}
\begin{split}
G^{r}(x,x',\omega)=
\begin{cases}
\Psi_{1}(x)[\alpha_{11}\tilde{\Psi}_{5}^{T}(x')+\alpha_{12}\tilde{\Psi}_{6}^{T}(x')+\alpha_{13}\tilde{\Psi}_{7}^{T}(x')+\alpha_{14}\tilde{\Psi}_{8}^{T}(x')]\\
+
\Psi_{2}(x)[\alpha_{21}\tilde{\Psi}_{5}^{T}(x')+\alpha_{22}\tilde{\Psi}_{6}^{T}(x')+\alpha_{23}\tilde{\Psi}_{7}^{T}(x')+\alpha_{24}\tilde{\Psi}_{8}^{T}(x')]\\
+
\Psi_{3}(x)[\alpha_{31}\tilde{\Psi}_{5}^{T}(x')+\alpha_{32}\tilde{\Psi}_{6}^{T}(x')+\alpha_{33}\tilde{\Psi}_{7}^{T}(x')+\alpha_{34}\tilde{\Psi}_{8}^{T}(x')]\\
+
\Psi_{4}(x)[\alpha_{41}\tilde{\Psi}_{5}^{T}(x')+\alpha_{42}\tilde{\Psi}_{6}^{T}(x')+\alpha_{43}\tilde{\Psi}_{7}^{T}(x')+\alpha_{44}\tilde{\Psi}_{8}^{T}(x')]
\,,\quad x>x'&\\
\Psi_{5}(x)[\beta_{11}\tilde{\Psi}_{1}^{T}(x')+\beta_{12}\tilde{\Psi}_{2}^{T}(x')+\beta_{13}\tilde{\Psi}_{3}^{T}(x')+\beta_{14}\tilde{\Psi}_{4}^{T}(x')]\\
+\Psi_{6}(x)[\beta_{21}\tilde{\Psi}_{1}^{T}(x')+\beta_{22}\tilde{\Psi}_{2}^{T}(x')+\beta_{23}\tilde{\Psi}_{3}^{T}(x')+\beta_{24}\tilde{\Psi}_{4}^{T}(x')]\\
+\Psi_{7}(x)[\beta_{31}\tilde{\Psi}_{1}^{T}(x')+\beta_{32}\tilde{\Psi}_{2}^{T}(x')+\beta_{33}\tilde{\Psi}_{3}^{T}(x')+\beta_{34}\tilde{\Psi}_{4}^{T}(x')]\\
+\Psi_{8}(x)[\beta_{41}\tilde{\Psi}_{1}^{T}(x')+\beta_{42}\tilde{\Psi}_{2}^{T}(x')+\beta_{43}\tilde{\Psi}_{3}^{T}(x')+\beta_{44}\tilde{\Psi}_{4}^{T}(x')]\,, \quad x<x'&
\end{cases}
\end{split}
\end{equation}
Coefficients $\alpha_{ij}$ and $\beta_{mn}$ in Eq.~\eqref{RGF} are calculated from
\begin{equation}
[\omega-H_{BdG}(x)]G^{r}(x,x',\omega)=\delta(x-x'),
\label{rgf1}
\end{equation}
where $H_{BdG}(x)=\begin{pmatrix}
H\hat{I} & i\Delta \Theta(x)\hat{\sigma}_{y} \\
-i\Delta^{*}\Theta(x)\hat{\sigma}_{y} & -H\hat{I}
\end{pmatrix},$ with $H=p^2/2m^\star-J_{0}\delta(x)\vec s.\vec S -E_{F}$. Eq.~\eqref{rgf1} on integration at $x=x'$ yields,
\begin{equation}
\label{conditionGRSO}
[G^{r}(x>x')]_{x=x'}=[G^{r}(x<x')]_{x=x'},\,\,\, \mbox{ and }\,\,\,
[\frac{d}{dx}G^{r}(x>x')]_{x=x'}-[\frac{d}{dx}G^{r}(x<x')]_{x=x'}=\eta\tau_{z}\sigma_{0},
\end{equation}
wherein $\eta=2m^{*}/\hbar^2$ and $\tau_{i}$, $\sigma_{i}$ are Pauli matrices in particle-hole and spin spaces. Green's functions is a $2\times2$ matrix in particle-hole space,
\begin{equation}
\label{GF}
G^{r}(x,x',\omega)=
\begin{bmatrix}
G^{r}_{ee}&G^{r}_{eh}\\
G^{r}_{he}&G^{r}_{hh}
\end{bmatrix},
\end{equation}
where each component of $G^{r}(x,x',\omega$) is a matrix. In presence of spin flip scattering, individual Green's function element can be written as
\begin{equation}
G_{\alpha\beta}^{r}(x,x',\omega)=
\begin{pmatrix}
[G^{r}_{\alpha\beta}]_{\uparrow\uparrow}&[G^{r}_{\alpha\beta}]_{\uparrow\downarrow}\\
[G^{r}_{\alpha\beta}]_{\downarrow\uparrow}&[G^{r}_{\alpha\beta}]_{\downarrow\downarrow}
\end{pmatrix}, \mbox{with } \alpha,\beta \in \{e,h\}.
\label{geh}
\end{equation}
Analytical expressions of all Green's functions are provided in Appendix A.
\subsection{Pairing amplitudes}
Spin symmetry of anomalous Green's function propagator is calculated from,
\begin{equation}
G^{r}_{eh}(x,x',\omega)=i\sum_{\lambda=0}^{3}f_{\lambda}^{r}\sigma_{\lambda}\sigma_{2},
\label{green}
\end{equation}
where $\sigma_{0}$ is the identity matrix, $\sigma_{\lambda}(\lambda=1,2,3)$ are Pauli matrices. In Eq.~\eqref{green}, $f_{0}^{r}$ is spin-singlet ($\uparrow\downarrow-\downarrow\uparrow$), $f_{1,2}^{r}$ are equal spin-triplet ($\downarrow\downarrow\pm\uparrow\uparrow$) and $f_{3}^{r}$ is mixed spin triplet ($\uparrow\downarrow+\downarrow\uparrow$) component. Equal spin triplet components $\uparrow\uparrow$ and $\downarrow\downarrow$ are given by $f_{\uparrow\uparrow}=if_{2}^{r}-f_{1}^{r}$ and $f_{\downarrow\downarrow}=if_{2}^{r}+f_{1}^{r}$, respectively. Also, $f_{\lambda}^{r}$ denotes pairing amplitude or correlation, while $|f_{\lambda}^{r}|=\sqrt{(f_{\lambda}^{r})(f_{\lambda}^{r})^{*}}$ refers to pairing magnitude. Using Eqs.~\eqref{GF},\eqref{geh} and Eq.~\eqref{green} we get pairing amplitudes or correlations as
\begin{equation}
\label{pairingfunctions}
\begin{split}
&f_{0}^{r}(x,x',\omega)= \frac{[G^{r}_{eh}]_{\uparrow\downarrow}-[G^{r}_{eh}]_{\downarrow\uparrow}}{2},\,\,
f_{1}^{r}(x,x',\omega)= \frac{[G^{r}_{eh}]_{\downarrow\downarrow}-[G^{r}_{eh}]_{\uparrow\uparrow}}{2},\,\,\\
&f_{2}^{r}(x,x',\omega)= \frac{[G^{r}_{eh}]_{\downarrow\downarrow}+[G^{r}_{eh}]_{\uparrow\uparrow}}{2i}, \mbox{ and }
f_{3}^{r}(x,x',\omega)= \frac{[G^{r}_{eh}]_{\uparrow\downarrow}+[G^{r}_{eh}]_{\downarrow\uparrow}}{2}\,.
\end{split}
\end{equation}
The even and odd frequency components can be extracted by using
\begin{equation}
\label{EVENODD}
f^{E}_{\lambda}(x,x',\omega)=\frac{f^{r}_{\lambda}(x,x',\omega)+f^{a}_{\lambda}(x,x',-\omega)}{2},\,\, \mbox{ and }\,\,
f^{O}_{\lambda}(x,x',\omega)=\frac{f^{r}_{\lambda}(x,x',\omega)-f^{a}_{\lambda}(x,x',-\omega)}{2},
\end{equation}
where $f_{\lambda}^{a}$ is related to advanced Green's function, which can be found from retarded Green's functions using\cite{cayy} $G^{a}(x,x',\omega)=[G^{r}(x',x,\omega)]^{\dagger}$. Even and odd frequency components of equal spin triplet correlations can be obtained from Eq.~\eqref{EVENODD},
\begin{equation}
\label{eo1}
f_{\uparrow\uparrow}^{E}=if_{2}^{E}-f_{1}^{E},\,\,\, f_{\downarrow\downarrow}^{E}=if_{2}^{E}+f_{1}^{E},\,\,\,
f_{\uparrow\uparrow}^{O}=if_{2}^{O}-f_{1}^{O},\,\,\, \mbox{ and }\,\,\, f_{\downarrow\downarrow}^{O}=if_{2}^{O}+f_{1}^{O}.
\end{equation}
At finite temperature we go to the Matsubara representation and replace $\omega$ with $ i\omega_{n}$. In this case Eq.~\eqref{green} can be written\cite{chan} as
\begin{equation}
\label{greenFT}
\sum_{\omega_{n}>0} G^{r}_{eh}(x,x',i\omega_{n})=i\sum_{\lambda=0}^{3}f_{\lambda}^{r}\sigma_{\lambda}\sigma_{2},
\end{equation}
$\omega_{n}=\pi k_{B}T(2n+1)$ are Matsubara frequencies and $n=0,\pm1,\pm2,...$.
\subsection{Spin polarized local density of states(SPLDOS)}
LDOS $\nu(x,\omega)$ and local magnetization density of state (LMDOS) $\vec{m}(x,\omega)$ can be calculated\cite{kuz} from retarded Green's function,
\begin{equation}
\label{lod}
\nu(x,\omega)=-\frac{1}{\pi}\lim_{\epsilon\rightarrow0}\text{Im}[\text{Tr}\{G^{r}_{ee}(x,x,\omega+i\epsilon)\}],\,\, \mbox{ and }\,\,
\textbf{m}(x,\omega)=-\frac{1}{\pi}\lim_{\epsilon\rightarrow0}\text{Im}[\text{Tr}\{\vec{\sigma_{\lambda}}.G^{r}_{ee}(x,x,\omega+i\epsilon)\}].
\end{equation}
From Eq.~\eqref{lod}, spin up ($\sigma=+1$) and spin-down ($\sigma=-1$) components of SPLDOS are calculated as $\nu_{\sigma}=(\nu+\sigma|\textbf{m}|)/2$.
\section{Analysis}
Following the procedure mentioned in the previous section, we analyze induced odd/even frequency spin-singlet and spin-triplet correlations in both N ($x<0$) and S ($x>0$) regions at zero and finite temperature and also discuss SPLDOS with particular focus on its relation with odd frequency pairing correlation.
\subsection{N-SF-S junction (Zero temperature)}
\subsubsection{Odd and even frequency spin singlet correlations}
Induced odd/even frequency pairing amplitudes or correlations are directly calculated from anomalous particle-hole component of retarded Green's function using Eqs.~\eqref{pairingfunctions}, \eqref{EVENODD} and \eqref{eo1}. Detailed derivation is provided in the Appendix A. For even and odd frequency spin singlet correlations we get,
\begin{eqnarray}
\label{singleteven}
&&f_{0}^{E}(x,x',\omega)=\begin{cases}-\frac{i\eta a_{12}}{2k_{e}}e^{-ik^{M}(x+x')}\cos[k_{F}(x-x')], \quad \mbox{for}\,\, x<0\,\, \mbox{(normal metal region)} \\
\frac{\eta u v}{2i(u^2-v^2)}e^{-\kappa|x-x'|}\Bigg[\frac{e^{ik_{F}|x-x'|}}{k_{e}^{S}}+\frac{e^{-ik_{F}|x-x'|}}{k_{h}^{S}}\Bigg]+\frac{\eta u v}{2i(u^2-v^2)}e^{-\kappa(x+x')}\Bigg[\frac{b_{51}e^{ik_{F}(x+x')}}{k_{e}^{S}}+\frac{b_{82}e^{-ik_{F}(x+x')}}{k_{h}^{S}}\Bigg]\\
+\frac{\eta }{2i(u^2-v^2)}e^{-\kappa(x+x')}\frac{a_{81}\cos[k_{F}(x-x')](k_{F}+i\kappa(u^2-v^2))}{(k_{F}^2+\kappa^2)}, \quad \mbox{for}\,\, x>0\,\, \mbox{(superconducting region)}
\end{cases}\\
\label{singletodd}
&&f_{0}^{O}(x,x',\omega)=\begin{cases}-\frac{\eta a_{12}}{2k_{e}}e^{-ik^{M}(x+x')}\sin[k_{F}(x-x')], \quad \mbox{for}\,\, x<0\,\, \mbox{(normal metal region)} \\
\frac{\eta a_{81}(k_{F}(u^2-v^2)+i\kappa)}{2(u^2-v^2)(k_{F}^2+\kappa^2)}\sin[k_{F}(x-x')]e^{-\kappa(x+x')}, \quad \mbox{for}\,\, x>0\,\, \mbox{(superconducting region)}
\end{cases}
\end{eqnarray}
where $k^{M}=\omega k_{F}/(2E_{F})$. Both even frequency spin singlet (ESE) and odd frequency spin singlet (OSO) correlations are interface contributions in normal metal(N) region ($x<0$) as is evident, being proportional to Andreev reflection amplitude $a_{12}$. In S region ($x>0$), ESE correlations have a bulk contribution (first term of Eq.~\eqref{singleteven} for $x>0$), in addition to interface contribution while OSO correlations have only an interface contribution. Bulk contribution to ESE correlations in S region ($x>0$), from Eq.~\ref{singleteven} (independent of interface scattering amplitudes) is
\begin{equation}
f_{0,B}^{E}=\frac{\eta u v}{2i(u^2-v^2)}e^{-\kappa|x-x'|}\Bigg[\frac{e^{ik_{F}|x-x'|}}{k_{e}^{S}}+\frac{e^{-ik_{F}|x-x'|}}{k_{h}^{S}}\Bigg],
\label{evenbulk}
\end{equation}
while interface contributions from Eq.~\eqref{singleteven} which depend on interface scattering amplitudes, are
\begin{equation}
f_{0,I}^{E}=\frac{\eta u v}{2i(u^2-v^2)}e^{-\kappa(x+x')}\Bigg[\frac{b_{51}e^{ik_{F}(x+x')}}{k_{e}^{S}}+\frac{b_{82}e^{-ik_{F}(x+x')}}{k_{h}^{S}}\Bigg]+\frac{\eta }{2i(u^2-v^2)}e^{-\kappa(x+x')}\frac{a_{81}\cos[k_{F}(x-x')](k_{F}+i\kappa(u^2-v^2))}{(k_{F}^2+\kappa^2)}.
\label{evenint}
\end{equation}
Bulk contribution does not exhibit any local space dependence, as substituting $x=x'$ in (Eq.~\eqref{evenbulk}) makes them independent of $x$. In contrast interface contribution (Eq.~\eqref{evenint}) is $x$ dependent. Thus, for $x\rightarrow \infty$ bulk contribution (Eq.~\eqref{evenbulk}) is finite, while interface contribution (Eq.~\eqref{evenint}) vanishes. Therefore, local even frequency spin singlet correlation is finite in bulk. From Eq.~\eqref{singleteven} we see that ESE correlation in S region depends on both normal reflection ($b_{51}$, $b_{82}$) and Andreev reflection amplitude ($a_{81}$), while in Eq.~\eqref{singletodd} OSO correlation in S region is proportional to Andreev reflection amplitude $a_{81}$. At $x=x'$, OSO vanishes, while ESE is finite and becomes maximum.
\begin{figure}[h]
\centering{\includegraphics[width=.99\textwidth]{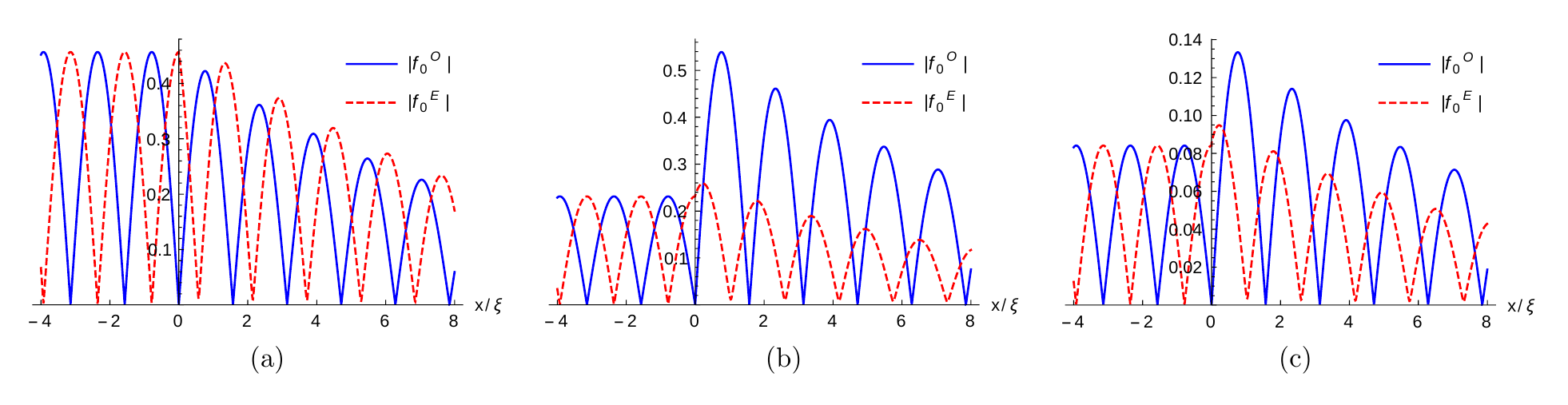}}
\caption{\small \sl Absolute values of even and odd frequency spin-singlet correlation induced in N ($x<0$) and S ($x>0$) regions as a function of position $x$ for (a) no flip case and (b,c) spin flip case. Parameters are: $S=1/2$ (for (a) and (b)), $S=5/2$ (for (c)), $F=F'=0$ (for (a)), $F=F'=1$ (for (b)), $F=F'=3$ (for (c)), $J=1$, $x'=0$, $\omega=0.1\Delta$, $E_{F}=10\Delta$.}
\end{figure}

We plot spin-singlet pairing correlation induced in N and S regions in Fig.~2. In Fig.~2(a) OSO (blue, solid) and ESE (red, dashed) correlations are finite and show nice periodic oscillations as function of position $x$ in N ($x<0$) region, while in S ($x>0$) region both OSO and ESE correlations exhibit an oscillatory decay for case of no spin-flip scattering (flip probability $F=F'=0$, see Eqs.~\eqref{eu}, \eqref{ed}). The decay length $1/\kappa$ with $\kappa=\sqrt{\Delta^2-\omega^2}[k_{F}/(2E_{F})]$, and oscillatory nature of correlations is determined from Fermi-energy $E_{F}$ (through $k_{F}$, see Eqs.~\eqref{singleteven}, \eqref{singletodd} for $x>0$). In Fig.~2(b) for finite spin-flip scattering (flip probability $F=F'=1$, see Eqs.~\eqref{eu}, \eqref{ed}), we see similar oscillatory behavior, albeit with lower magnitude than no flip case. The more interesting thing is shown in Fig.~2(b) in the S region, wherein we see OSO correlation dominates over ESE correlation. For high values of spin-flip scattering ($F=F'=3$), both OSO and ESE correlations are suppressed, but their nature does not change, as shown in Fig.~2(c). Further, we also note that non-local ($x\neq x'$) spin-singlet correlations are finite regardless of odd or even frequency. In contrast, local ($x=x'$) OSO correlations vanish, and ESE correlations are finite. In all the figures, $x$ has been normalized by superconducting coherence length $\xi$ to make it dimensionless.
\subsubsection{Odd and even frequency equal spin-triplet correlations}
Even and odd frequency, spin-triplet correlation can be of two distinct types, mixed spin-triplet or equal spin triplet. To distinguish between these two, we denote odd frequency equal spin-triplet correlation as (OTE-equal) while odd frequency mixed spin-triplet correlation as (OTE-mixed). Similarly, in the case of even frequency, we have ETO-equal and ETO-mixed correlations. Uniquely, we find that mixed spin-triplet correlation for both odd and even frequency vanishes ($f_{3}^{E}=f_{3}^{O}=0$) for our set-up (Fig.~1), while equal spin-triplet correlation for both odd and even frequency is finite ($f_{\uparrow\uparrow}^{E,O}=-f_{\downarrow\downarrow}^{E,O}\neq0$).

The reason why mixed spin triplet correlations (both even and odd frequency) vanish, can be traced to the scattering amplitudes and Green's functions derived from them in sections II. B and III. If one looks at the Andreev reflection amplitudes, they satisfy the relations: $a_{31}=-a_{42}$ and $a_{32}=-a_{41}$. Thus, anomalous electron-hole components of Green's functions in N region are related as: $[G^{r}_{eh}]_{\uparrow\uparrow}=-[G^{r}_{eh}]_{\downarrow\downarrow}$ and $[G^{r}_{eh}]_{\uparrow\downarrow}=-[G^{r}_{eh}]_{\downarrow\uparrow}$ (see Appendix A for detailed calculation). This same relation, for anomalous electron-hole components of Green's functions holds true in S($x>0$) region as well. Therefore, from Eq.~\eqref{pairingfunctions}, $f_{2}^{r}=0$ and $f_{3}^{r}=0$, with $f_{\uparrow\uparrow}=-f_{\downarrow\downarrow}$. Spin flip scattering at junction interfaces, induces equal spin triplet correlations only.
Interestingly, we find odd frequency equal spin-triplet correlation (OTE-equal) dominating over even frequency equal spin-triplet correlation (ETO-equal) in the S region, which can have significant applications in superconducting spintronics.

For the even/odd frequency equal spin triplet correlations, using Eq.~\eqref{eo1} we obtain
\begin{eqnarray}
\label{tripleteven}
&&f_{\uparrow\uparrow}^{E}(x,x',\omega)=-f_{\downarrow\downarrow}^{E}(x,x',\omega)=\begin{cases}-\frac{\eta a_{11}}{2k_{e}}e^{-ik^{M}(x+x')}\sin[k_{F}(x-x')], \quad \mbox{for}\,\, x<0\,\, \mbox{(normal metal(N) region)} \\
-\frac{\eta a_{62}(k_{F}(u^2-v^2)+i\kappa)}{2(u^2-v^2)(k_{F}^2+\kappa^2)}\sin[k_{F}(x-x')]e^{-\kappa(x+x')}, \quad \mbox{for}\,\, x>0\,\, \mbox{(superconducting(S) region)}
\end{cases}\\
\label{tripletodd}
&&f_{\uparrow\uparrow}^{O}(x,x',\omega)=-f_{\downarrow\downarrow}^{O}(x,x',\omega)=\begin{cases}-\frac{i\eta a_{11}}{2k_{e}}e^{-ik^{M}(x+x')}\cos[k_{F}(x-x')], \quad \mbox{for}\,\, x<0\,\, \mbox{(N region)} \\
\frac{\eta u v}{2i(u^2-v^2)}e^{-\kappa(x+x')}\Bigg[\frac{b_{72}e^{-ik_{F}(x+x')}}{k_{h}^{S}}-\frac{b_{61}e^{ik_{F}(x+x')}}{k_{e}^{S}}\Bigg]\\-\frac{\eta}{2i(u^2-v^2)}e^{-\kappa(x+x')}\frac{a_{62}\cos[k_{F}(x-x')](k_{F}+i\kappa(u^2-v^2))}{(k_{F}^2+\kappa^2)}, \quad \mbox{for}\,\, x>0\,\, \mbox{(S region)}
\end{cases}
\end{eqnarray}
in presence of spin flip scattering, while in absence of spin flip scattering they completely vanish. In absence of spin flip scattering, both incident quasi particle spin and spin flipper's spin are in same direction (either up or down) and they do not flip their spins after interaction. For $S=1/2$, $m'$ can be $m'=1/2$ (when spin flipper's spin is in up direction) or $m'=-1/2$ (when spin flipper's spin is in down direction). {When spin up quasiparticle is incident, no-flip process implies $S=m'$ (i.e., $F=0$) while when spin down quasiparticle is incident $S=-m'$ (i.e., $F'=0$) . }
When spin up quasiparticle is incident and $m'=1/2$, then there wont be spin flip scattering ($F=0$) and from Eq.~\eqref{eu} we get
\begin{equation}
\label{eun}
\vec s.\vec S \varphi_{1}^{N}\phi_{\frac{1}{2}}^{\frac{1}{2}}=\frac{1}{4}\varphi_{1}^{N}\phi_{\frac{1}{2}}^{\frac{1}{2}}.
\end{equation}
Similarly, when spin down quasiparticle is incident and $m'=-1/2$, then there will not be any spin flip scattering ($F'=0$) and from Eq.~\eqref{ed} we get
\begin{equation}
\label{edn}
\vec s.\vec S \varphi_{2}^{N}\phi_{-\frac{1}{2}}^{\frac{1}{2}}=\frac{1}{4}\varphi_{2}^{N}\phi_{-\frac{1}{2}}^{\frac{1}{2}}.
\end{equation}
From Eqs.~\eqref{eun}, \eqref{edn} we see that in absence of spin-flip scattering $\vec{s}.\vec{S}$ operates similarly on spin up and spin down quasiparticle spinors, and therefore the system becomes spin-inactive.
Thus, at NS interface when spin flipper does not flip its spin, spin-singlet to spin-triplet conversion can not take place, and spin triplet correlations do not arise in such a situation (see Appendix A for detailed calculation of how spin triplet correlations vanish in absence of spin flip scattering). From Eqs.~\eqref{tripleteven}, \eqref{tripletodd}, we see that both even and odd frequency, equal spin triplet correlations are interface contributions. We further notice that in S($x>0$) region ETO-equal correlation is proportional to Andreev reflection amplitude $a_{62}$, while OTE-equal correlation depends on both normal and Andreev reflection amplitudes. In N ($x<0$) region both even and odd frequency equal spin triplet correlations are proportional to Andreev reflection amplitude $a_{11}$. At $x=x'$, local ETO-equal correlations vanish, i.e., $f_{\uparrow\uparrow}^{E}=-f_{\downarrow\downarrow}^{E}=0$, but local OTE-equal correlation is finite, i.e, $f_{\uparrow\uparrow}^{O}=-f_{\downarrow\downarrow}^{O}\neq0$.

In Fig.~3 both ETO-equal and OTE-equal correlations are shown as a function of position $x$ for low ($F=F'=1$, Fig.~3(a)) and high ($F=F'=3$, Fig.~3(b)) values of spin flip scattering. We see that in the metallic region $f_{\uparrow\uparrow}^{E,O}$ or $f_{\downarrow\downarrow}^{E,O}$ is finite and exhibits an oscillatory behavior as function of position $x$ and survives infinitely far away in presence of spin flip scattering.
\begin{figure}[h]
\centering{\includegraphics[width=.99\textwidth]{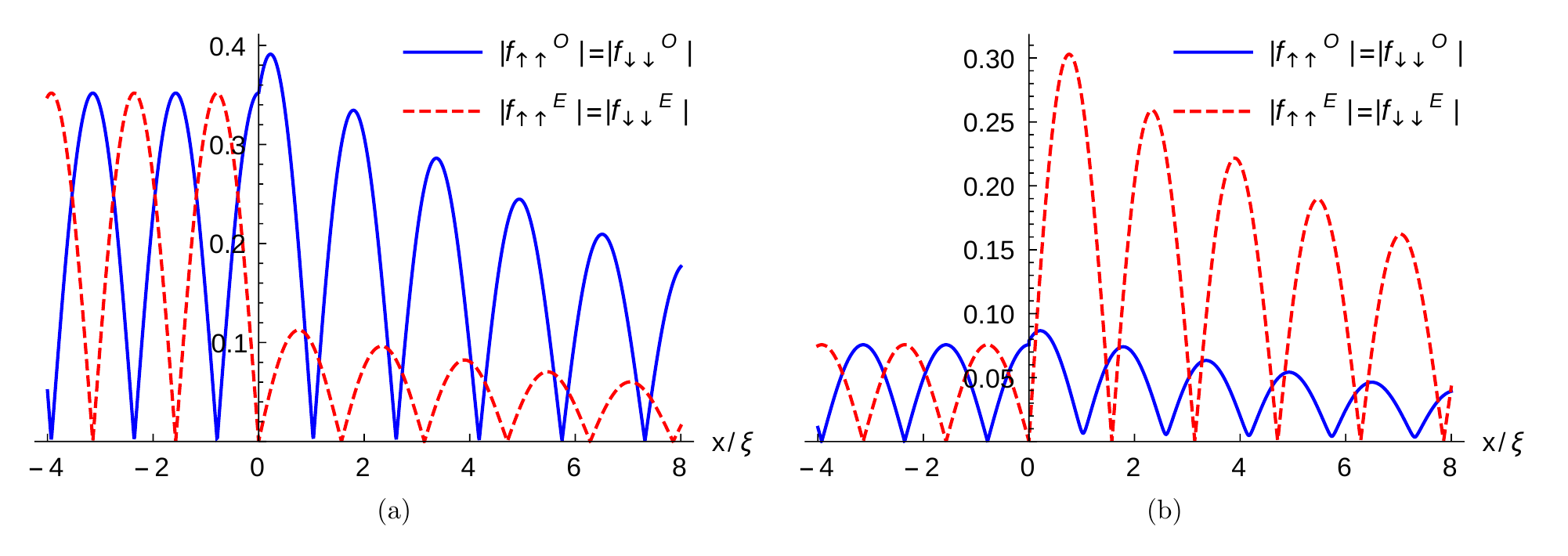}}
\caption{\small \sl Absolute values of even and odd frequency equal spin-triplet correlation induced in N ($x<0$) and S ($x>0$) regions as function of position $x$ for spin flip case. Parameters are: $S=1/2$ (for (a)), $S=5/2$ (for (b)), $F=F'=1$ (for (a)), $F=F'=3$ (for (b)), $J=1$, $x'=0$, $\omega=0.1\Delta$, $E_{F}=10\Delta$.}
\end{figure}
The reason for this kind of behavior can be understood from Eqs.~\eqref{tripleteven}, \eqref{tripletodd} for $x<0$, where we see that the equal spin-triplet pairings are proportional to $\sin[k_{F}(x-x')]$ for even frequency and $\cos[k_{F}(x-x')]$ for odd frequency respectively. However, in the S region, we find that in the presence of spin-flip scattering, both ETO-equal and OTE-equal correlations exhibit an oscillatory decay, in contrast to what we see for spin-triplet correlations in the normal metal region ($x<0$). One can understand the different results for equal spin-triplet correlations in N and S regions from Eq.~\eqref{tripleteven} for $x>0$, where we see that equal spin-triplet correlation in S region is proportional to $\sin[k_{F}(x-x')]e^{-\kappa(x+x')}$ and therefore shows an oscillatory decay with decay length $1/\kappa$. In contrast, in the N region, correlations don't decay. Another interesting thing to note from Fig.~3 is that in a ballistic NS junction, non-local even and odd frequency equal spin-triplet correlations are finite only in the presence of spin-flip scattering, which is a remarkable result of our paper. Further, at $x=x'$, we notice that ETO-equal correlations vanish, but OTE-equal correlations are non-zero. In addition, we notice that in the S region, the OTE-equal correlation is larger than the ETO-equal correlation for low values of spin-flip scattering. In contrast, for high values of spin-flip scattering, the ETO-equal correlation dominates over the OTE-equal correlation. Finally, we reiterate that both ETO-mixed and OTE-mixed correlations vanish regardless of spin-flip scattering, i.e., $f_{3}^{E}=f_{3}^{O}=0$ in our setup.

As an aside, for an NS junction based on 1D nanowires with Rashba spin-orbit coupling~\cite{amb} and proximity-induced, $s$-wave spin-singlet superconductivity, even/odd frequency mixed spin-triplet correlations are induced with vanishing even/odd frequency equal spin-triplet correlations. Spin-orbit coupling does not generate equal spin-triplet correlations in NS junctions.
\subsubsection{ Spin polarized local density of states (SPLDOS)}
After analyzing spin triplet correlations, we now discuss SPLDOS in both N($x<0$) and S($x>0$) regions and try to find any relationship with odd frequency pairing. For SPLDOS, from Eq.~\eqref{lod} we obtain
\begin{eqnarray}
\label{ldns}
\nu_{\sigma}(x,\omega)=\begin{cases}\frac{1}{2\pi}\text{Im}\Bigg[\frac{i\eta(1+b_{11}e^{-i2k_{e}x})}{k_{e}}\Bigg]+\frac{\sigma}{2\pi}\sqrt{\text{Im}\Bigg[\frac{i\eta b_{12}e^{-i2k_{e}x}}{k_{e}}}\Bigg]^2, \quad \mbox{for}\,\, x<0\,\, \mbox{(normal metal region)} \\
e^{-2\kappa x}\times\Bigg[\frac{1}{2\pi}\text{Im}[\rho_{1}]+\frac{\sigma}{2\pi}\sqrt{\text{Im}[\rho_{2}]^2}\Bigg], \quad \mbox{for}\,\, x>0\,\, \mbox{(superconducting region)}
\end{cases}
\end{eqnarray}
\begin{eqnarray}
\mbox{ where, } \rho_{1}=&&\frac{i\eta(2a_{81}k_{F}uv+b_{51}e^{i2k_{F}x}u^2(k_{F}-i\kappa)+b_{82}e^{-i2k_{F}x}v^2(k_{F}+i\kappa)+e^{2\kappa x}(k_{F}-i(u^2-v^2)\kappa))}{(u^2-v^2)(k_{F}^2+\kappa^2)},\nonumber\\
\rho_{2}=&&\frac{i\eta(b_{72}v^2(k_{F}+i\kappa)e^{-i2k_{F}x}-2a_{62}k_{F}uv-b_{61}u^2(k_{F}-i\kappa)e^{i2k_{F}x})}{(u^2-v^2)(k_{F}^2+\kappa^2)}.\nonumber
\end{eqnarray}
From Eq.~\eqref{ldns} we see that SPLDOS has a bulk as well as interface contribution. Further, there is a decay term $e^{-2\kappa x}$ in Eq.~\eqref{ldns} for $x>0$. The first term in Eq.~\eqref{ldns} represents LDOS, while the second term represents LMDOS. For LMDOS, from Eq.~\eqref{lod} we get
\begin{eqnarray}
\label{lmdos}
\textbf{m}(x,\omega)=\begin{cases}\frac{1}{\pi}\text{Im}\Big[\frac{i\eta b_{12}e^{-i2k_{e}x}}{k_{e}}\Big]\hat{x}, \quad \mbox{for}\,\, x<0\,\, \mbox{(normal metal region)} \\
\frac{1}{\pi}\text{Im}[\rho_{2}]\hat{x}, \quad \mbox{for}\,\, x>0\,\, \mbox{(superconducting region)}
\end{cases}
\end{eqnarray}
From Eq.~\eqref{lmdos} we see that LMDOS is parallel to $x$ axis. The reason why LMDOS is parallel to $x$ direction, can be found to the scattering amplitudes and Green's functions obtained from them in sections II. B and III. If one looks at the normal reflection amplitudes, they satisfy the relations: $b_{11}=b_{22}$ and $b_{12}=b_{21}$. Thus, normal electron-electron components of Green's functions in N region are related as: $[G^{r}_{ee}]_{\uparrow\uparrow}=[G^{r}_{ee}]_{\downarrow\downarrow}$ and $[G^{r}_{ee}]_{\uparrow\downarrow}=[G^{r}_{ee}]_{\downarrow\uparrow}$ (see Appendix A for detailed calculation). This same relation, for normal electron-electron components of Green's functions holds true in S region as well. Therefore, from Eq.~\eqref{lod}, $y$ and $z$ components of LMDOS vanish and LMDOS is parallel to $x$ axis.
In presence of spin-flip scattering, scattering amplitude $b_{12}$ (normal metal region) and $\rho_{2}$ (in superconducting region) are finite in Eq.~\eqref{ldns}. Thus, spin up and spin down components of SPLDOS are not equal, i.e., $\nu_{\uparrow}\neq\nu_{\downarrow}$. Thus, in presence of spin-flip scattering LDOS is spin-polarized, although $f_{\uparrow\uparrow}=-f_{\downarrow\downarrow}$. However, in absence of spin-flip scattering both $b_{12}$ and $\rho_{2}$ are zero in Eq.~\eqref{ldns}. Thus, in absence of spin-flip scattering, LDOS is not spin-polarized, i.e., $\nu_{\uparrow}=\nu_{\downarrow}=\nu$. While, in Ref.~\cite{amb}, the presence of Rashba spin-orbit coupling leads to mixed spin-triplet correlations and spin-polarized LDOS. In our work, spin-flip scattering leads to equal spin-triplet correlations and spin-polarized LDOS. However, in Ref.~\cite{amb}, in the absence of spin-orbit coupling, as also in our work in the absence of spin-flip scattering, LDOS is not spin-polarized. To conclude, while in Ref.~\onlinecite{amb} spin-orbit coupling is responsible for spin polarization of LDOS, in our work, spin-flip scattering is responsible for spin polarization of LDOS.

In Figs.~4(a) and 5(a) we plot spin-up and spin-down LDOS as a function of $\omega$ at the NS interface ($x=0$). We see that for low values of spin flip scattering ($F=F'=1$) when $f_{\uparrow\uparrow}^{O}$ dominates over $f_{\uparrow\uparrow}^{E}$, there is a peak at $\omega=0$. But, for increasing values of spin flip scattering ($F=F'=3$) when even frequency-equal spin triplet correlation (ETO-equal), i.e., $f_{\uparrow\uparrow}^{E}$ is greater than odd frequency equal spin triplet correlation (OTE-equal), i.e., $f_{\uparrow\uparrow}^{O}$, there is a dip at $\omega=0$ in Fig.~5(a). In Figs.~4(b) and 5(b) we plot spin polarized LDOS as function of position $x$ in both N($x<0$) and S($x>0$) regions. We notice that SPLDOS in N region shows nice periodic oscillations, while SPLDOS in S region ($x>0$) exhibits an oscillatory decay due to normal reflection.
\begin{figure}[h]
\centering{\includegraphics[width=.99\textwidth]{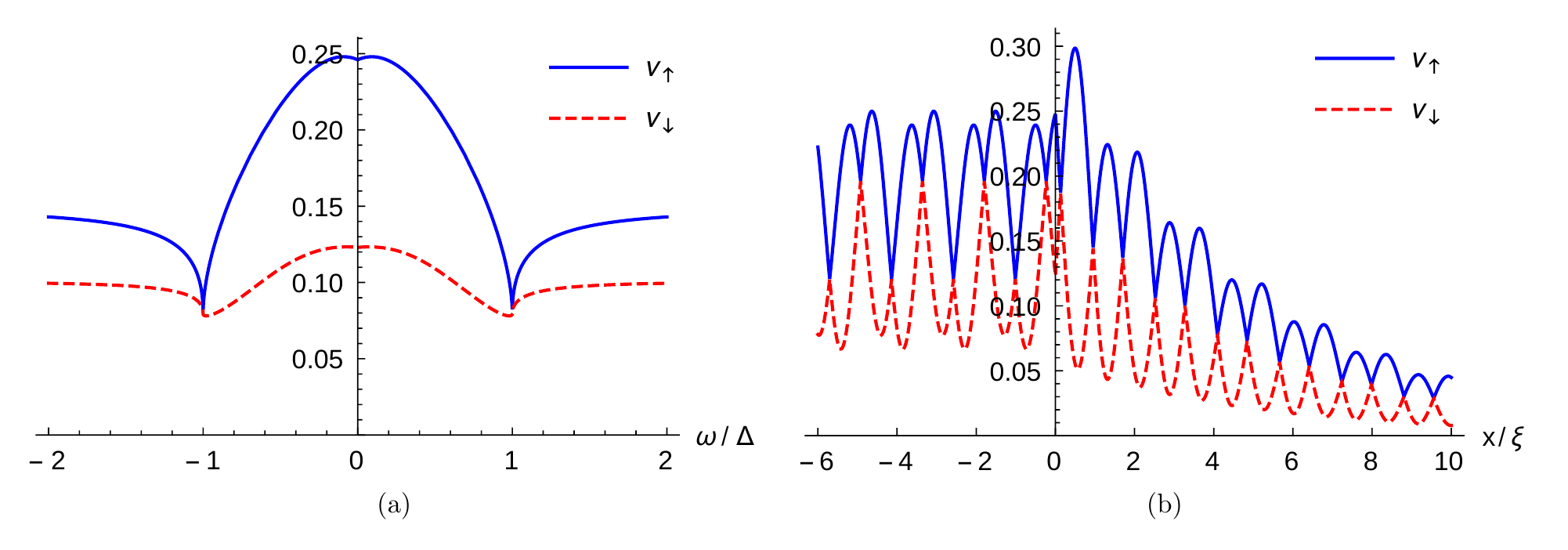}}
\caption{\small \sl (a) Frequency dependence of SPLDOS at NS interface, (b) spatial dependence of the SPLDOS in N($x<0$) and S($x>0$) regions. Parameters are: $S=1/2$, $F=F'=1$, $J=1$, $x=0$ (for (a)), $E_{F}=10\Delta$, $\omega=0.1\Delta$ (for (b)).}
\end{figure}
\begin{figure}[h]
\centering{\includegraphics[width=.99\textwidth]{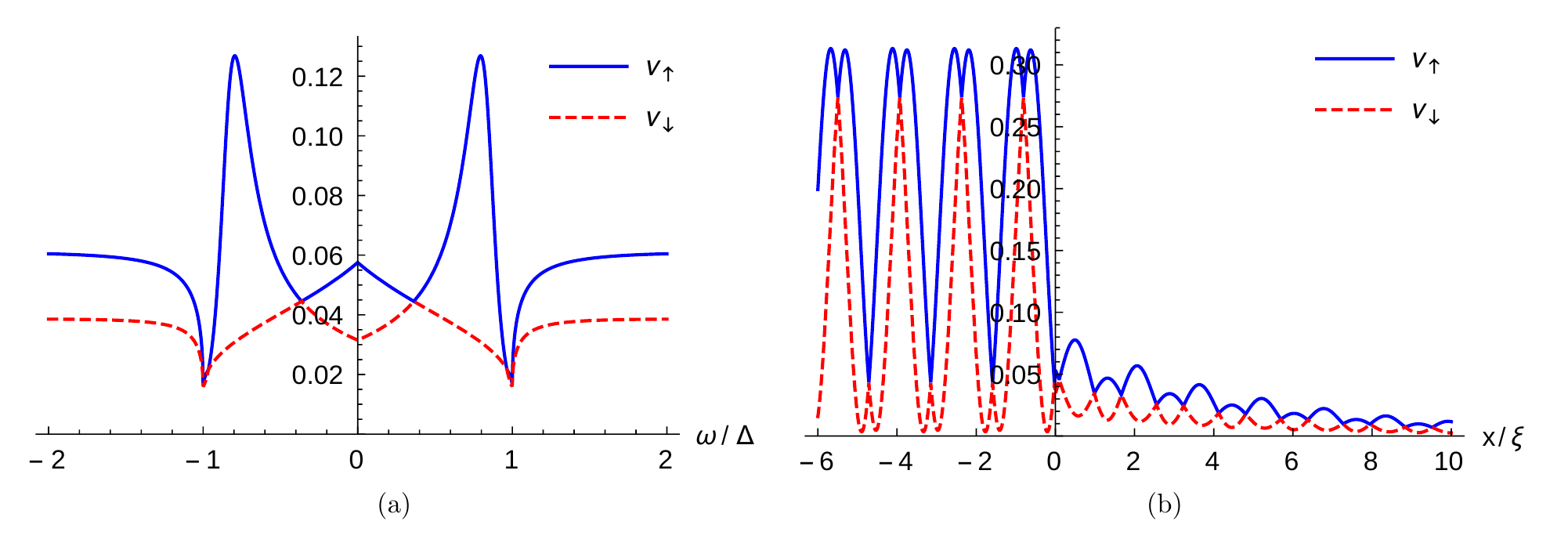}}
\caption{\small \sl (a) Frequency dependence of SPLDOS at NS interface, (b) spatial dependence of the SPLDOS in N($x<0$) and S($x>0$) regions. Parameters are: $S=5/2$, $F=F'=3$, $J=1$, $x=0$ (for (a)), $E_{F}=10\Delta$, $\omega=0.1\Delta$ (for (b)).}
\end{figure}
Next we analyze, possible relation between odd and even frequency pairing amplitudes with SPLDOS. SPLDOS in N region depends on normal reflection (see Eq.~\eqref{ldns} for $x<0$), while even and odd frequency spin singlet as well as spin triplet correlation in N region depends only on Andreev reflection (see Eqs.~\eqref{singleteven}, \eqref{singletodd}, \eqref{tripleteven}, \eqref{tripletodd} for $x<0$). In S region, SPLDOS depends on both normal and Andreev reflection (see Eq.~\eqref{ldns} for $x>0$), while only even frequency spin singlet correlation and odd frequency equal spin triplet correlation in S region depend on both normal and Andreev reflection (see Eqs.~\eqref{singleteven}, \eqref{tripletodd} for $x>0$). In addition, since SPLDOS is local measurement, it is quite natural to analyze only local ($x=x'$) odd and even frequency correlations. We see that only ESE and OTE-equal correlations survive, see Eqs.~\eqref{localsinglet} and \eqref{localtriplet} below,
\begin{eqnarray}
\label{localsinglet}
&&f_{0}^{E,L}(x,\omega)=\begin{cases}-\frac{i\eta a_{12}}{2k_{e}}e^{-2ik^{M}x}, \quad \mbox{for}\,\, x<0\,\, \mbox{(normal metal region)} \\
\frac{\eta u v}{2i(u^2-v^2)}\Bigg[\frac{1}{k_{e}^{S}}+\frac{1}{k_{h}^{S}}\Bigg]+\frac{\eta u v}{2i(u^2-v^2)}e^{-2\kappa x}\Bigg[\frac{b_{51}e^{2ik_{F}x}}{k_{e}^{S}}+\frac{b_{82}e^{-2ik_{F}x}}{k_{h}^{S}}\Bigg]\\
+\frac{\eta }{2i(u^2-v^2)}e^{-2\kappa x}\frac{a_{81}(k_{F}+i\kappa(u^2-v^2))}{(k_{F}^2+\kappa^2)}, \quad \quad \mbox{ for}\,\, x>0\,\, \mbox{(superconducting region)}
\end{cases}\\
\label{localtriplet}
&&f_{\uparrow\uparrow}^{O,L}(x,\omega)=-f_{\downarrow\downarrow}^{O,L}(x,\omega)=\begin{cases}-\frac{i\eta a_{11}}{2k_{e}}e^{-2ik^{M}x}, \quad \mbox{for}\,\, x<0\,\, \mbox{(normal metal region)}\\
e^{-2\kappa x}\times\Bigg[\frac{\eta u v}{2i(u^2-v^2)}\Bigg(\frac{b_{72}e^{-i2k_{F}x}}{k_{h}^{S}}-\frac{b_{61}e^{i2k_{F}x}}{k_{e}^{S}}\Bigg)-\frac{\eta}{2i(u^2-v^2)}\frac{a_{62}(k_{F}+i\kappa(u^2-v^2))}{(k_{F}^2+\kappa^2)}\Bigg], \quad \\ \mbox{for}\,\, x>0\,\, \mbox{(superconducting region)}.
\end{cases}
\end{eqnarray}
Since OSO and ETO-equal correlations are sine functions as shown in Eqs.~\eqref{singletodd}, \eqref{tripleteven}, they vanish. In Fig.~6, we present spatial dependence of local odd and even frequency correlations and spin-up LDOS in the N($x < 0$) and S($x>0$) regions. In the S region ($x>0$), both local odd frequency equal spin-triplet correlation and spin-up LDOS show an exponential decay and nice oscillatory behavior. But, in the N region, $x<0$, local odd and even frequency correlations are independent of position $x$, while spin-up LDOS exhibits a nice periodic oscillation.
\begin{figure}[h]
{\includegraphics[width=0.99\textwidth]{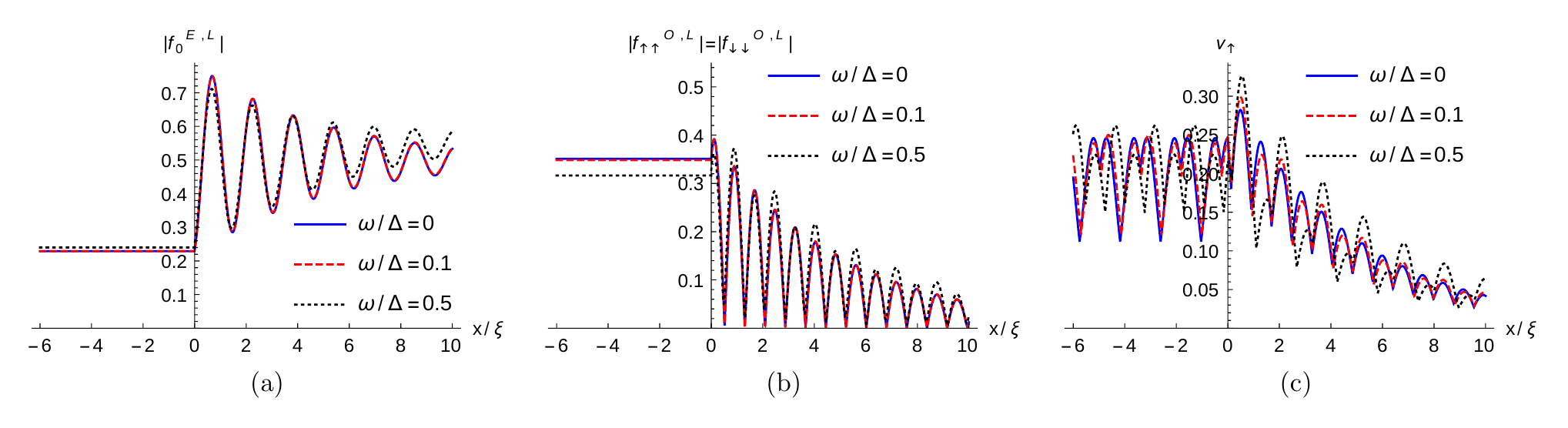}}
\caption{\small \sl Spatial dependence of the (a) local even frequency correlations, (b) local odd frequency correlations and (c) spin-up LDOS in N and S regions. Parameters are: $S=1/2$, $F=F'=1$, $J=1$, $E_{F}=10\Delta$.}
\end{figure}
Figs.~6(a) and 6(b) show that only local OTE-equal correlations show a nice oscillatory decay in the S region similar to spin-up LDOS. In contrast, local ESE correlations exhibit an oscillatory behavior without decay. Therefore, one is justified in associating only local odd frequency equal spin-triplet correlation with SPLDOS. By computing SPLDOS, we can extract the associated coefficients and calculate odd frequency equal spin-triplet correlations in the superconducting region of our junction. Consequently, a large SPLDOS indicates significant odd frequency equal spin-triplet pairing whose signature can be seen experimentally. SPLDOS has been used to detect odd frequency correlations, e.g., in Ref.~\cite{kuz} wherein a Kondo-type magnetic impurity is embedded in a $s$-wave superconductor. Odd frequency mixed spin-triplet correlations dominate at the critical value of magnetic impurity strength. The SPLDOS shows the same spatial and frequency behavior as the odd frequency mixed spin-triplet correlations. Local odd frequency correlations show the same frequency and spatial dependence as the LMDOS, which can be detected experimentally using spin-polarized scanning tunneling spectroscopy. The effect of finite temperature on spin-singlet and triplet correlations is discussed in Appendix B.

{
\section{Processes at play}
In this section, we explain the reasons behind our results. We examine three different situations: (a) when both spin-flip scattering and spin mixing are present in the system, (b) when only spin mixing is present in the system, and (c) when only spin-flip scattering is present in the system. Spin mixing and spin-flip scattering are two separate processes. In the spin mixing process, an electron experiences spin-dependent phase shifts\cite{lind}, while in the spin-flip scattering process, an electron flips its spin\cite{AJP}. For example, when an electron propagates through a ferromagnetic layer, only spin mixing occurs\cite{dutta}.
On the other hand, when an electron propagates through a ferromagnetic bilayer with misaligned magnetization, both spin mixing and spin-flip scattering occur\cite{hme,toll,daum}. However, only spin-flip scattering occurs when an electron interacts with the spin flipper. We will discuss below the spin structure of the retarded Green's functions and induced pairing correlations and SPLDOS for each of these three cases.
\subsection{Retarded Green's functions and induced pairing correlations}
\subsubsection{Both spin mixing and spin-flip scattering occur}
In the case of Superconductor-Ferromagnet-Ferromagnet-Superconductor (S-F$_{1}$-F$_{2}$-S) junction or Ferromagnet-Ferromagnet-Superconductor (F$_{1}$-F$_{2}$-S) junction, with misaligned magnetizations, both spin mixing and spin-flip scattering occur when an electron/hole propagates through the ferromagnetic bilayer. The S-F$_{1}$-F$_{2}$-S junction case has been dealt with elaborately in Ref.~\cite{hme}. Hereinbelow we show the calculations for F$_{1}$F$_{2}$S junction.}
{
We consider an one dimensional Ferromagnet (F$_1$)-Ferromagnet (F$_{2}$)-Superconductor (S) junction as shown in Fig.~7, where the magnetization vectors of the two Ferromagnets make an angle $\theta$ with each other. In Fig.~7, the scattering of an up spin electron incident is shown and normal reflection, Andreev reflection and quasi-particle transmission into superconductor are represented. The model Hamiltonian in BdG formalism of the system as depicted in Fig.~7 is given as:
\begin{equation}
H_{F, BdG}(x)=
\begin{pmatrix}
H_{F}\hat{I} & i\Delta \Theta(x)\hat{\sigma}_{y} \\
-i\Delta^{*}\Theta(x)\hat{\sigma}_{y} & -H_{F}\hat{I}
\end{pmatrix},
\end{equation}
where $H_{F}=p^2/2m^\star-\vec{h_{1}}.\hat{\sigma}\Theta(-x-a)-\vec{h_{2}}.\hat{\sigma}[\Theta(x+a)+\Theta(-x)]-E_{F}$. The magnetization vector ($\vec{h_{2}}$) of right ferromagnetic layer ($F_{2}$) is at an angle $\theta$ with $z$ axis in the $y-z$ plane, while that of left ferromagnetic layer ($F_{1}$) is fixed along the $z$ axis. Thus, $\vec{h_{2}}.\hat{\sigma}=h_{2}\sin \theta\hat{\sigma}_{y}+h_{2}\cos \theta\hat{\sigma}_{z}$.
\begin{figure}[h]
\centering{\includegraphics[width=.99\textwidth]{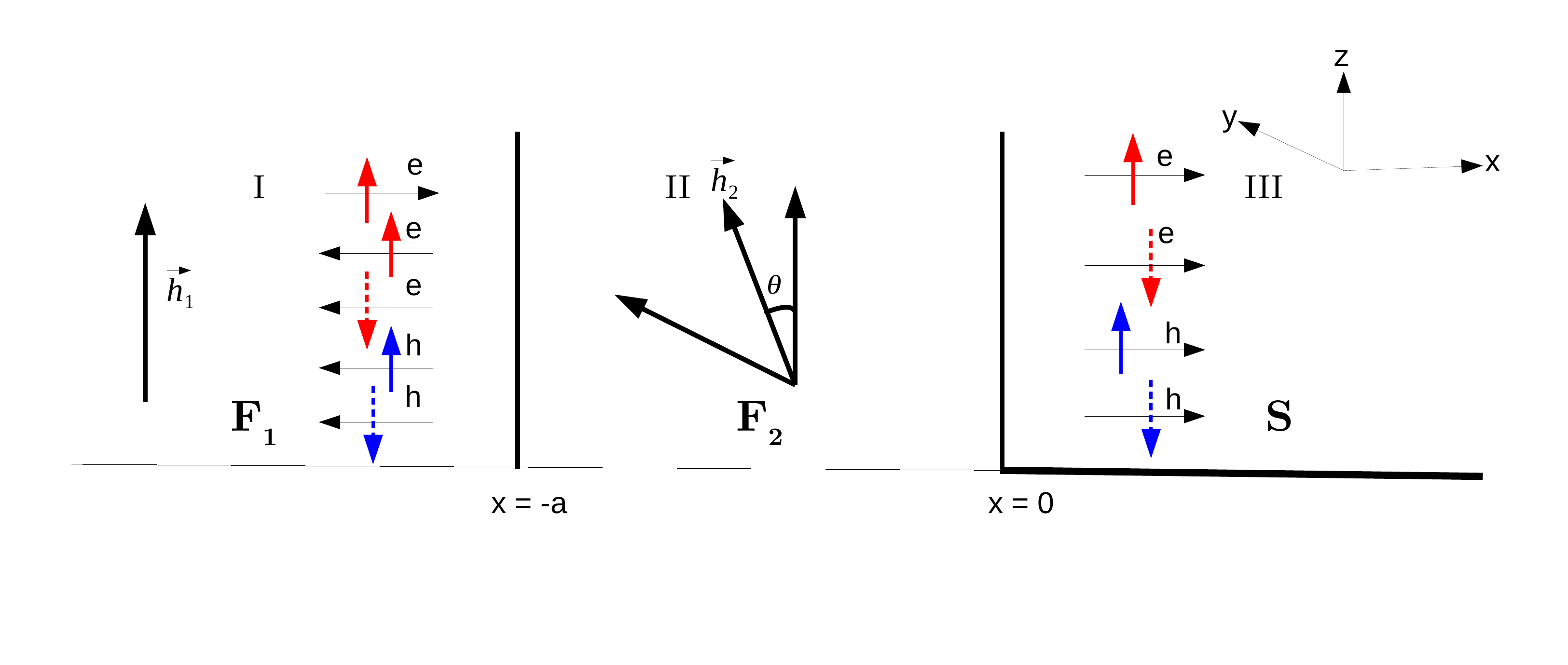}}
\caption{\small \sl {Ferromagnet (F$_{1}$)-Ferromagnet (F$_{2}$)-Superconductor (S) junction with misaligned magnetizations. The scattering of a spin up electron incident is shown. Normal reflection, Andreev reflection and quasi-particle transmission into superconductor are depicted.}}
\end{figure}
If we diagonalize BdG Hamiltonian $H_{F, BdG}(x)$ we will get wavefunctions in different regions of our system for various types of scattering processes. Wavefunctions for different types of scattering processes are given as-
\begin{equation}
\begin{split}
\Psi_{1}(x)&=\begin{cases}
\varphi_{1}^{N}e^{iq_{\uparrow}^{+}(x+a)}+a'_{11}\varphi_{3}^{N}e^{iq_{\uparrow}^{-}(x+a)}+a'_{12}\varphi_{4}^{N}e^{iq_{\downarrow}^{-}(x+a)}+b'_{11}\varphi_{1}^{N}e^{-iq_{\uparrow}^{+}(x+a)}+b'_{12}\varphi_{2}^{N}e^{-iq_{\downarrow}^{+}(x+a)}\,,& x<-a\\
c'_{11}\varphi_{1}^{F}e^{iq_{\uparrow}^{+}(x+a)}+c'_{12}\varphi_{2}^{F}e^{iq_{\downarrow}^{+}(x+a)}+d'_{11}\varphi_{1}^{F}e^{-iq_{\uparrow}^{+}x}+d'_{12}\varphi_{2}^{F}e^{-iq_{\downarrow}^{+}x}+e'_{11}\varphi_{3}^{F}e^{iq_{\uparrow}^{-}x}+e'_{12}\varphi_{4}^{F}e^{iq_{\downarrow}^{-}x}\\
+f'_{11}\varphi_{3}^{F}e^{-iq_{\uparrow}^{-}(x+a)}+f'_{12}\varphi_{4}^{F}e^{-iq_{\downarrow}^{-}(x+a)}\,,& -a<x<0\\
g'_{11}\varphi_{1}^{S}e^{ik_{e}^{S}x}+g'_{12}\varphi_{2}^{S}e^{ik_{e}^{S}x}+h'_{11}\varphi_{3}^{S}e^{-ik_{h}^{S}x}+h'_{12}\varphi_{4}^{S}e^{-ik_{h}^{S}x}\,,& x>0
\end{cases}\\
\Psi_{2}(x)&=\begin{cases}
\varphi_{2}^{N}e^{iq_{\downarrow}^{+}(x+a)}+a'_{21}\varphi_{3}^{N}e^{iq_{\uparrow}^{-}(x+a)}+a'_{22}\varphi_{4}^{N}e^{iq_{\downarrow}^{-}(x+a)}+b'_{21}\varphi_{1}^{N}e^{-iq_{\uparrow}^{+}(x+a)}+b'_{22}\varphi_{2}^{N}e^{-iq_{\downarrow}^{+}(x+a)}\,,& x<-a\\
c'_{21}\varphi_{1}^{F}e^{iq_{\uparrow}^{+}(x+a)}+c'_{22}\varphi_{2}^{F}e^{iq_{\downarrow}^{+}(x+a)}+d'_{21}\varphi_{1}^{F}e^{-iq_{\uparrow}^{+}x}+d'_{22}\varphi_{2}^{F}e^{-iq_{\downarrow}^{+}x}+e'_{21}\varphi_{3}^{F}e^{iq_{\uparrow}^{-}x}+e'_{22}\varphi_{4}^{F}e^{iq_{\downarrow}^{-}x}\\
+f'_{21}\varphi_{3}^{F}e^{-iq_{\uparrow}^{-}(x+a)}+f'_{22}\varphi_{4}^{F}e^{-iq_{\downarrow}^{-}(x+a)}\,,& -a<x<0\\
g'_{21}\varphi_{1}^{S}e^{ik_{e}^{S}x}+g'_{22}\varphi_{2}^{S}e^{ik_{e}^{S}x}+h'_{21}\varphi_{3}^{S}e^{-ik_{h}^{S}x}+h'_{22}\varphi_{4}^{S}e^{-ik_{h}^{S}x}\,,& x>0
\end{cases}\\
\Psi_{3}(x)&=\begin{cases}
\varphi_{3}^{N}e^{-iq_{\uparrow}^{-}(x+a)}+a'_{31}\varphi_{1}^{N}e^{-iq_{\uparrow}^{+}(x+a)}+a'_{32}\varphi_{2}^{N}e^{-iq_{\downarrow}^{+}(x+a)}+b'_{31}\varphi_{3}^{N}e^{iq_{\uparrow}^{-}(x+a)}+b'_{32}\varphi_{4}^{N}e^{iq_{\downarrow}^{-}(x+a)}\,,& x<-a\\
c'_{31}\varphi_{1}^{F}e^{iq_{\uparrow}^{+}(x+a)}+c'_{32}\varphi_{2}^{F}e^{iq_{\downarrow}^{+}(x+a)}+d'_{31}\varphi_{1}^{F}e^{-iq_{\uparrow}^{+}x}+d'_{32}\varphi_{2}^{F}e^{-iq_{\downarrow}^{+}x}+e'_{31}\varphi_{3}^{F}e^{iq_{\uparrow}^{-}x}+e'_{32}\varphi_{4}^{F}e^{iq_{\downarrow}^{-}x}\\
+f'_{31}\varphi_{3}^{F}e^{-iq_{\uparrow}^{-}(x+a)}+f'_{32}\varphi_{4}^{F}e^{-iq_{\downarrow}^{-}(x+a)}\,,& -a<x<0\\
g'_{31}\varphi_{1}^{S}e^{ik_{e}^{S}x}+g'_{32}\varphi_{2}^{S}e^{ik_{e}^{S}x}+h'_{31}\varphi_{3}^{S}e^{-ik_{h}^{S}x}+h'_{32}\varphi_{4}^{S}e^{-ik_{h}^{S}x}\,,& x>0
\end{cases}\\
\Psi_{4}(x)&=\begin{cases}
\varphi_{4}^{N}e^{-iq_{\downarrow}^{-}(x+a)}+a'_{41}\varphi_{1}^{N}e^{-iq_{\uparrow}^{+}(x+a)}+a'_{42}\varphi_{2}^{N}e^{-iq_{\downarrow}^{+}(x+a)}+b'_{41}\varphi_{3}^{N}e^{iq_{\uparrow}^{-}(x+a)}+b'_{42}\varphi_{4}^{N}e^{iq_{\downarrow}^{-}(x+a)}\,,& x<-a\\
c'_{41}\varphi_{1}^{F}e^{iq_{\uparrow}^{+}(x+a)}+c'_{42}\varphi_{2}^{F}e^{iq_{\downarrow}^{+}(x+a)}+d'_{41}\varphi_{1}^{F}e^{-iq_{\uparrow}^{+}x}+d'_{42}\varphi_{2}^{F}e^{-iq_{\downarrow}^{+}x}+e'_{41}\varphi_{3}^{F}e^{iq_{\uparrow}^{-}x}+e'_{42}\varphi_{4}^{F}e^{iq_{\downarrow}^{-}x}\\
+f'_{41}\varphi_{3}^{F}e^{-iq_{\uparrow}^{-}(x+a)}+f'_{42}\varphi_{4}^{F}e^{-iq_{\downarrow}^{-}(x+a)}\,,& -a<x<0\\
g'_{41}\varphi_{1}^{S}e^{ik_{e}^{S}x}+g'_{42}\varphi_{2}^{S}e^{ik_{e}^{S}x}+h'_{41}\varphi_{3}^{S}e^{-ik_{h}^{S}x}+h'_{42}\varphi_{4}^{S}e^{-ik_{h}^{S}x}\,,& x>0
\end{cases}\\
\Psi_{5}(x)&=\begin{cases}
g'_{51}\varphi_{1}^{N}e^{-iq_{\uparrow}^{+}(x+a)}+g'_{52}\varphi_{2}^{N}e^{-iq_{\downarrow}^{+}(x+a)}+h'_{51}\varphi_{3}^{N}e^{iq_{\uparrow}^{-}(x+a)}+h'_{52}\varphi_{4}^{N}e^{iq_{\downarrow}^{-}(x+a)}\,,& x<-a\\
c'_{51}\varphi_{1}^{F}e^{iq_{\uparrow}^{+}(x+a)}+c'_{52}\varphi_{2}^{F}e^{iq_{\downarrow}^{+}(x+a)}+d'_{51}\varphi_{1}^{F}e^{-iq_{\uparrow}^{+}x}+d'_{52}\varphi_{2}^{F}e^{-iq_{\downarrow}^{+}x}+e'_{51}\varphi_{3}^{F}e^{iq_{\uparrow}^{-}x}+e'_{52}\varphi_{4}^{F}e^{iq_{\downarrow}^{-}x}\\
+f'_{51}\varphi_{3}^{F}e^{-iq_{\uparrow}^{-}(x+a)}+f'_{52}\varphi_{4}^{F}e^{-iq_{\downarrow}^{-}(x+a)}\,,& -a<x<0\\
\varphi_{1}^{S}e^{-ik_{e}^{S}x}+a'_{51}\varphi_{3}^{S}e^{-ik_{h}^{S}x}+a'_{52}\varphi_{4}^{S}e^{-ik_{h}^{S}x}+b'_{51}\varphi_{1}^{S}e^{ik_{e}^{S}x}+b'_{52}\varphi_{2}^{S}e^{ik_{e}^{S}x}\,,& x>0
\end{cases}\\
\Psi_{6}(x)&=\begin{cases}
g'_{61}\varphi_{1}^{N}e^{-iq_{\uparrow}^{+}(x+a)}+g'_{62}\varphi_{2}^{N}e^{-iq_{\downarrow}^{+}(x+a)}+h'_{61}\varphi_{3}^{N}e^{iq_{\uparrow}^{-}(x+a)}+h'_{62}\varphi_{4}^{N}e^{iq_{\downarrow}^{-}(x+a)}\,,& x<-a\\
c'_{61}\varphi_{1}^{F}e^{iq_{\uparrow}^{+}(x+a)}+c'_{62}\varphi_{2}^{F}e^{iq_{\downarrow}^{+}(x+a)}+d'_{61}\varphi_{1}^{F}e^{-iq_{\uparrow}^{+}x}+d'_{62}\varphi_{2}^{F}e^{-iq_{\downarrow}^{+}x}+e'_{61}\varphi_{3}^{F}e^{iq_{\uparrow}^{-}x}+e'_{62}\varphi_{4}^{F}e^{iq_{\downarrow}^{-}x}\\
+f'_{61}\varphi_{3}^{F}e^{-iq_{\uparrow}^{-}(x+a)}+f'_{62}\varphi_{4}^{F}e^{-iq_{\downarrow}^{-}(x+a)}\,,& -a<x<0\\
\varphi_{2}^{S}e^{-ik_{e}^{S}x}+a'_{61}\varphi_{3}^{S}e^{-ik_{h}^{S}x}+a'_{62}\varphi_{4}^{S}e^{-ik_{h}^{S}x}+b'_{61}\varphi_{1}^{S}e^{ik_{e}^{S}x}+b'_{62}\varphi_{2}^{S}e^{ik_{e}^{S}x}\,,& x>0
\end{cases}\\
\Psi_{7}(x)&=\begin{cases}
g'_{71}\varphi_{1}^{N}e^{-iq_{\uparrow}^{+}(x+a)}+g'_{72}\varphi_{2}^{N}e^{-iq_{\downarrow}^{+}(x+a)}+h'_{71}\varphi_{3}^{N}e^{iq_{\uparrow}^{-}(x+a)}+h'_{72}\varphi_{4}^{N}e^{iq_{\downarrow}^{-}(x+a)}\,,& x<-a\\
c'_{71}\varphi_{1}^{F}e^{iq_{\uparrow}^{+}(x+a)}+c'_{72}\varphi_{2}^{F}e^{iq_{\downarrow}^{+}(x+a)}+d'_{71}\varphi_{1}^{F}e^{-iq_{\uparrow}^{+}x}+d'_{72}\varphi_{2}^{F}e^{-iq_{\downarrow}^{+}x}+e'_{71}\varphi_{3}^{F}e^{iq_{\uparrow}^{-}x}+e'_{72}\varphi_{4}^{F}e^{iq_{\downarrow}^{-}x}\\
+f'_{71}\varphi_{3}^{F}e^{-iq_{\uparrow}^{-}(x+a)}+f'_{72}\varphi_{4}^{F}e^{-iq_{\downarrow}^{-}(x+a)}\,,& -a<x<0\\
\varphi_{3}^{S}e^{ik_{h}^{S}x}+a'_{71}\varphi_{1}^{S}e^{ik_{e}^{S}x}+a'_{72}\varphi_{2}^{S}e^{ik_{e}^{S}x}+b'_{71}\varphi_{3}^{S}e^{-ik_{h}^{S}x}+b'_{72}\varphi_{4}^{S}e^{-ik_{h}^{S}x}\,,& x>0
\end{cases}\\
\Psi_{8}(x)&=\begin{cases}
g'_{81}\varphi_{1}^{N}e^{-iq_{\uparrow}^{+}(x+a)}+g'_{82}\varphi_{2}^{N}e^{-iq_{\downarrow}^{+}(x+a)}+h'_{81}\varphi_{3}^{N}e^{iq_{\uparrow}^{-}(x+a)}+h'_{82}\varphi_{4}^{N}e^{iq_{\downarrow}^{-}(x+a)}\,,& x<-a\\
c'_{81}\varphi_{1}^{F}e^{iq_{\uparrow}^{+}(x+a)}+c'_{82}\varphi_{2}^{F}e^{iq_{\downarrow}^{+}(x+a)}+d'_{81}\varphi_{1}^{F}e^{-iq_{\uparrow}^{+}x}+d'_{82}\varphi_{2}^{F}e^{-iq_{\downarrow}^{+}x}+e'_{81}\varphi_{3}^{F}e^{iq_{\uparrow}^{-}x}+e'_{82}\varphi_{4}^{F}e^{iq_{\downarrow}^{-}x}\\
+f'_{81}\varphi_{3}^{F}e^{-iq_{\uparrow}^{-}(x+a)}+f'_{82}\varphi_{4}^{F}e^{-iq_{\downarrow}^{-}(x+a)}\,,& -a<x<0\\
\varphi_{4}^{S}e^{ik_{h}^{S}x}+a'_{81}\varphi_{1}^{S}e^{ik_{e}^{S}x}+a'_{82}\varphi_{2}^{S}e^{ik_{e}^{S}x}+b'_{81}\varphi_{3}^{S}e^{-ik_{h}^{S}x}+b'_{82}\varphi_{4}^{S}e^{-ik_{h}^{S}x}\,,& x>0
\end{cases}\\
\end{split}
\label{wavfero}
\end{equation}
where $\varphi_{1}^{F}=\begin{pmatrix}
\cos\frac{\theta}{2}\\
i\sin\frac{\theta}{2}\\
0\\
0
\end{pmatrix}$, $\varphi_{2}^{F}=\begin{pmatrix}
i\sin\frac{\theta}{2}\\
\cos\frac{\theta}{2}\\
0\\
0
\end{pmatrix}$, $\varphi_{3}^{F}=\begin{pmatrix}
0\\
0\\
\cos\frac{\theta}{2}\\
-i\sin\frac{\theta}{2}
\end{pmatrix}$, $\varphi_{4}^{F}=\begin{pmatrix}
0\\
0\\
-i\sin\frac{\theta}{2}\\
\cos\frac{\theta}{2}
\end{pmatrix}$.
$\Psi_{1}$, $\Psi_{2}$, $\Psi_{3}$ and $\Psi_{4}$ represents the scattering processes when spin up electron, spin down electron, spin up hole and spin down hole are incident from ferromagnetic region I respectively, while $\Psi_{5}$, $\Psi_{6}$, $\Psi_{7}$ and $\Psi_{8}$ represents the scattering processes when spin up electron, spin down electron, spin up hole and spin down hole are incident from superconducting region respectively. $b'_{ij}$ and $a'_{ij}$ are normal reflection amplitudes and Andreev reflection amplitudes respectively, while $g'_{ij}$ and $h'_{ ij}$ are transmission amplitudes for electron-like quasi-particles and hole-like quasi-particles respectively. $q_{\sigma}^{\pm}=\sqrt{\frac{2m^{*}}{\hbar^2}(E_{F}\pm\omega+\rho_{\sigma}h)}$ are the wave-vectors for electron ($q_{\sigma}^{+}$) and hole ($q_{\sigma}^{-}$) in the Ferromagnet, with $\rho_{\sigma}=+1(-1)$ when- $\sigma=\uparrow(\downarrow)$. Conjugated processes $\tilde{\psi_{i}}$ needed to construct the Green's functions are determined by diagonalizing the Hamiltonian $H_{F,BdG}^{*}(-k)$ instead of $H_{F,BdG}(k)$. In case of Ferromagnet-Ferromagnet-Superconductor junction (Fig.~7) we find that $\tilde{\varphi_{i}}^{N(S)}=\varphi_{i}^{N(S)}$ and $\tilde{\varphi_{i}}^{F}=(\varphi_{i}^{F})^{*}$. In the limit of $E_{F}>>\Delta,\omega$ we approximate $q_{\sigma}^{\pm}\approx k_{F}(1\pm\frac{\omega}{2E_{F}}+\rho_{\sigma}\frac{h}{2E_{F}})$ with $k_{F}=\sqrt{2m^{*}E_{F}/\hbar^2}$.
Scattering amplitudes are obtained from the boundary conditions. Boundary condition at $x=-a$ is-
\begin{eqnarray}
\label{BCF1}
{}&\psi_{i}(x<-a)=\psi_{i}(-a<x<0),\\
&\mbox { and, }\frac{d\psi_{i}(-a<x<0)}{dx}-\frac{d\psi_{i}(x<-a)}{dx}=0.
\end{eqnarray}
Boundary condition at $x=0$ is-
\begin{eqnarray}
{}&\psi_{i}(-a<x<0)=\psi_{i}(x>0),\\
&\mbox{ and, }\frac{d\psi_{i}(x>0)}{dx}-\frac{d\psi_{i}(-a<x<0)}{dx}=0.
\label{BCF2}
\end{eqnarray}
Solving the above boundary conditions, we get 16 equations for each type of scattering process as discussed in Eq.~\eqref{wavfero}.
From each set of these 16 equations we can determine the different scattering amplitudes. Using these scattering amplitudes and following the similar procedure as discussed in section III, we can compute retarded Green's function and induced pairing correlations in each region
of junction. {Detailed calculations are shown in Appendix C.}
For even and odd frequency spin singlet correlations, using Eq.~\eqref{EVENODD} we get,
\begin{equation}
\begin{split}
&f_{0}^{E}(x,x',\omega)=\frac{i\eta}{4}\Bigg[\frac{a'_{32}e^{-ik^{N}(x+x')}}{q_{\uparrow}^{-}}-\frac{a'_{41}e^{-ik^{N'}(x+x')}}{q_{\downarrow}^{-}}\Bigg]\cos[k_{F}(x-x')],\,\,\mbox{and}\\
&f_{0}^{O}(x,x',\omega)=\frac{\eta}{4}\Bigg[\frac{a'_{32}e^{-ik^{N}(x+x')}}{q_{\uparrow}^{-}}-\frac{a'_{41}e^{-ik^{N'}(x+x')}}{q_{\downarrow}^{-}}\Bigg]\sin[k_{F}(x-x')],
\end{split}
\end{equation}
where $k^{N}=\frac{(\omega-h_{1})k_{F}}{2E_{F}}$ and $k^{N'}=\frac{(\omega+h_{1})k_{F}}{2E_{F}}$. Similarly, for even and odd frequency equal spin triplet correlations, using Eqs.~\eqref{EVENODD} and \eqref{eo1} we obtain,
\begin{equation}
\begin{split}
&f_{\uparrow\uparrow}^{E}(x,x',\omega)=-\frac{\eta a'_{31}}{2q_{\uparrow}^{-}}e^{-ik^{M}(x+x')}\sin[k^{L}(x-x')],\,\,
f_{\uparrow\uparrow}^{O}(x,x',\omega)=-\frac{i\eta a'_{31}}{2q_{\uparrow}^{-}}e^{-ik^{M}(x+x')}\cos[k^{L}(x-x')],\\
&f_{\downarrow\downarrow}^{E}(x,x',\omega)=-\frac{\eta a'_{42}}{2q_{\downarrow}^{-}}e^{-ik^{M}(x+x')}\sin[k^{L'}(x-x')], \mbox{ and }
f_{\downarrow\downarrow}^{O}(x,x',\omega)=-\frac{i\eta a'_{42}}{2q_{\downarrow}^{-}}e^{-ik^{M}(x+x')}\cos[k^{L'}(x-x')],
\end{split}
\label{equfer}
\end{equation}
where $k^{M}=\frac{\omega k_{F}}{2E_{F}}$, $k^{L}=k_{F}(1+\frac{h}{2E_{F}})$ and $k^{L'}=k_{F}(1-\frac{h}{2E_{F}})$. Finally, even and odd frequency mixed spin triplet correlations, using Eq.~\eqref{EVENODD} we get,
\begin{equation}
\begin{split}
&f_{3}^{E}(x,x',\omega)=-\frac{\eta}{4}\Bigg[\frac{a'_{32}e^{-ik^{N}(x+x')}}{q_{\uparrow}^{-}}-\frac{a'_{41}e^{-ik^{N'}(x+x')}}{q_{\downarrow}^{-}}\Bigg]\sin[k_{F}(x-x')],\,\,\mbox{and}\\
&f_{3}^{O}(x,x',\omega)=-\frac{i\eta}{4}\Bigg[\frac{a'_{32}e^{-ik^{N}(x+x')}}{q_{\uparrow}^{-}}-\frac{a'_{41}e^{-ik^{N'}(x+x')}}{q_{\downarrow}^{-}}\Bigg]\cos[k_{F}(x-x')].
\end{split}
\label{mixfer}
\end{equation}
From Eqs.~\eqref{equfer}, \eqref{mixfer} we see that both even and odd frequency equal and mixed spin-triplet correlations are finite when spin mixing and spin-flip scattering both are present in the system.
\subsubsection{Only spin mixing occurs}
In Fig.~7, when magnetization vectors of the two Ferromagnets are parallel to each other, i.e., $\theta=0$, then spin flip scattering does not occur and only spin mixing occurs in the system due to the exchange field of the Ferromagnets. Spin-mixing arises also in NS junction with Rashba spin-orbit coupling, see Ref.~\cite{amb}. In case of spin mixing process occuring in a FS junction (this is same as a F$_{1}$F$_{2}$S junction with aligned magnetization, and with vanishing length of F$_2$ layer, see Fig.~7), normal and Andreev reflection amplitudes with flip are zero, i.e., $b'_{12}=b'_{21}=a'_{31}=a'_{42}=0$. Thus, from Eq.~\eqref{grefer} {in Appendix C} we get $[G^{r}_{ee}]_{\uparrow\downarrow}=[G^{r}_{ee}]_{\downarrow\uparrow}=[G^{r}_{eh}]_{\uparrow\uparrow}=[G^{r}_{eh}]_{\downarrow\downarrow}=0$ and from Eq.~\eqref{equfer} we get $f_{\uparrow\uparrow}^{E}=f_{\uparrow\uparrow}^{O}=f_{\downarrow\downarrow}^{E}=f_{\downarrow\downarrow}^{O}=0$. Therefore, when only spin mixing occurs, even and odd frequency equal spin-triplet correlations vanish, but mixed spin-triplet correlations ($f_{3}^{r}$) are finite (see Eq.~\eqref{mixfer}).}
\subsubsection{Only spin flip scattering occurs}
In the case of Normal metal (N)-Spin flipper (SF)-Superconductor (S) junction, our chosen system depicted in Fig.~1, only spin-flip scattering occurs. In section IV.A.2, we have already shown that when only spin-flip scattering occurs, even and odd frequency equal spin-triplet correlations are finite, but mixed spin-triplet correlations vanish.

{In our work, there is a spin flipper at the N-S interface. When an electron/hole with spin up/down is incident from the metallic region at N-S interface, it interacts with the spin flipper through the exchange potential ($J_{0}\vec{s}.\vec{S}$), which may induce a mutual spin-flip. It results in electron/hole reflection into the N region with spin up or down and transmission of electron-like and hole-like quasiparticles with spin up or down into the S region for energies above the gap. Electron/hole does not experience any spin-dependent phase shifts when interacting with the spin flipper. Thus, there is no spin mixing, and only spin-flip scattering occurs. Spin flip scattering induces only equal spin-triplet correlations, as shown in Fig.~3.
\subsection{Spin polarized local density of states (SPLDOS) \& local magnetization density of states (LMDOS)}
\subsubsection{Both spin mixing and spin flip scattering occur}
In case of F$_{1}$F$_{2}$S junction with misaligned magnetizations as shown in Fig.~7, both spin mixing and spin flip scattering are present in the system. LMDOS for F$_{1}$F$_{2}$S junction, using Eq.~\eqref{lod} is given by,
\begin{equation}
\textbf{m}(x,\omega)=-\frac{1}{\pi}\lim_{\epsilon\rightarrow0}\text{Im}[([G^{r}_{ee}]_{\uparrow\downarrow}+[G^{r}_{ee}]_{\downarrow\uparrow})\hat{x}+i([G^{r}_{ee}]_{\uparrow\downarrow}-[G^{r}_{ee}]_{\downarrow\uparrow})\hat{y}+([G^{r}_{ee}]_{\uparrow\uparrow}-[G^{r}_{ee}]_{\downarrow\downarrow})\hat{z}],
\end{equation}
and SPLDOS is given by,
\begin{align}
\begin{split}
\nu_{\sigma}={}&-\frac{1}{2\pi}\lim_{\epsilon\rightarrow0}\text{Im}[([G^{r}_{ee}]_{\uparrow\uparrow}+[G^{r}_{ee}]_{\downarrow\downarrow})]\\{}&+\frac{\sigma}{2\pi}\lim_{\epsilon\rightarrow0}\sqrt{\text{Im}[([G^{r}_{ee}]_{\uparrow\downarrow}+[G^{r}_{ee}]_{\downarrow\uparrow})]^2+\text{Im}[i([G^{r}_{ee}]_{\uparrow\downarrow}-[G^{r}_{ee}]_{\downarrow\uparrow})]^2+\text{Im}[([G^{r}_{ee}]_{\uparrow\uparrow}-[G^{r}_{ee}]_{\downarrow\downarrow})]^2},
\end{split}
\end{align}
where $[G^{r}_{ee}]_{\uparrow\uparrow}$, $[G^{r}_{ee}]_{\uparrow\downarrow}$, $[G^{r}_{ee}]_{\downarrow\uparrow}$, and $[G^{r}_{ee}]_{\downarrow\downarrow}$ are mentioned in Eq.~\eqref{grefer} for left ferromagnetic region. Thus, both equal ($\uparrow\uparrow$ or $\downarrow\downarrow$) and mixed ($\uparrow\downarrow$ or $\downarrow\uparrow$) spin components of Green's function are finite and contribute to LMDOS and SPLDOS.}

{In Figs.~8(a) and 8(b) we plot even and odd frequency equal and mixed spin-triplet correlations respectively as a function of position $x$ in the superconducting region ($x>0$) for F$_{1}$F$_{2}$S junction when magnetization vectors of the two ferromagnetic layers are misaligned ($\theta\neq0$). We see that alongwith odd frequency correlations dominating over even frequency correlations, there is also a peak at $\omega=0$ for spin-up LDOS ($\nu_{\uparrow}$), while there is a dip at $\omega=0$ for spin-down LDOS ($\nu_{\downarrow}$) , see Fig.~8(c).}
\begin{figure}[h]
\centering{\includegraphics[width=.99\textwidth]{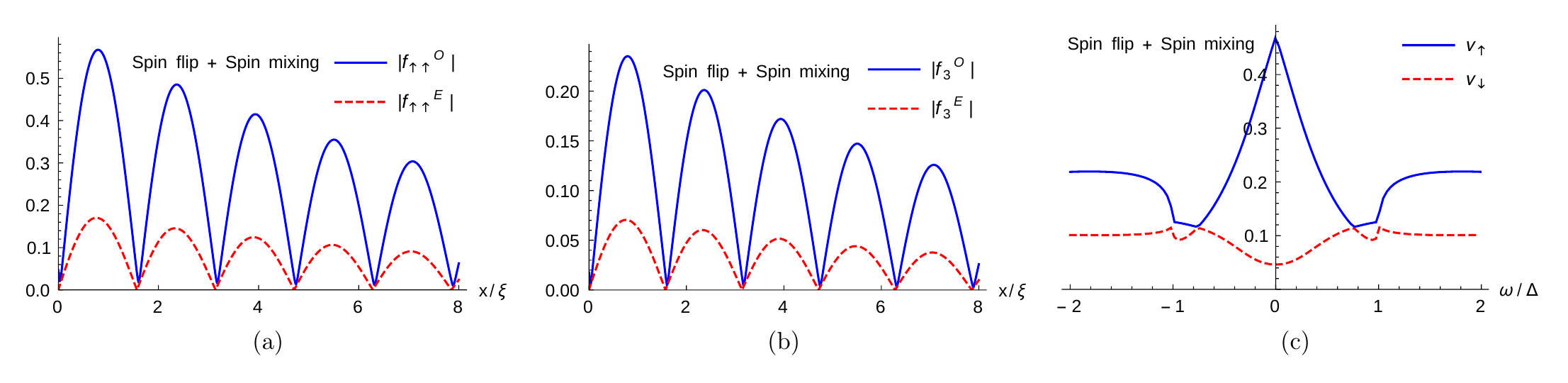}}
\caption{\small \sl{Absolute values of even and odd frequency (a) equal spin-triplet correlations and (b) mixed spin-triplet correlations induced in superconducting region for F$_{1}$F$_{2}$S junction as function of position $x$ when both spin flip scattering and spin mixing occur, (c) Frequency dependence of SPLDOS at $x=0$ for F$_{1}$F$_{2}$S junction when both spin flip scattering and spin mixing occur. Parameters are: $h_{1}/E_{F}=h_{2}/E_{F}=0.8$, $x'=0$, $\omega=0.1\Delta$ (for (a) and (b)), $E_{F}=10\Delta$, $\theta=\pi/2$, $k_{F}a=\pi$.}}
\end{figure}
{
\subsubsection{Only spin mixing occurs}
In case of a NS junction with Rashba spin-orbit coupling or a NFS junction or for a FS junction, spin flip scattering is absent and only spin mixing occurs. In absence of spin flip scattering, $[G^{r}_{ee}]_{\uparrow\downarrow}=[G^{r}_{ee}]_{\downarrow\uparrow}=0$. From the definition of LMDOS and SPLDOS, see Eq.~\eqref{lod}, one can calculate LMDOS and SPLDOS for a NFS junction or FS junction as,
\begin{equation}
\textbf{m}(x,\omega)=-\frac{1}{\pi}\lim_{\epsilon\rightarrow0}\text{Im}[([G^{r}_{ee}]_{\uparrow\uparrow}-[G^{r}_{ee}]_{\downarrow\downarrow})]\hat{z},
\end{equation}
and SPLDOS is given by,
\begin{equation}
\nu_{\sigma}=-\frac{1}{2\pi}\lim_{\epsilon\rightarrow0}\text{Im}[([G^{r}_{ee}]_{\uparrow\uparrow}+[G^{r}_{ee}]_{\downarrow\downarrow})]+\frac{\sigma}{2\pi}\lim_{\epsilon\rightarrow0}\sqrt{\text{Im}[([G^{r}_{ee}]_{\uparrow\uparrow}-[G^{r}_{ee}]_{\downarrow\downarrow})]^2}.
\end{equation}
Thus, only equal ($\uparrow\uparrow$ or $\downarrow\downarrow$) spin components of Green's function are finite and contribute to the LMDOS and SPLDOS.}

{
Since, there is no spin flip scattering, even and odd frequency equal spin-triplet correlations vanish and therefore in Fig.~9 we plot even and odd frequency mixed spin-triplet correlations and SPLDOS.
\begin{figure}[h]
\centering{\includegraphics[width=.99\textwidth]{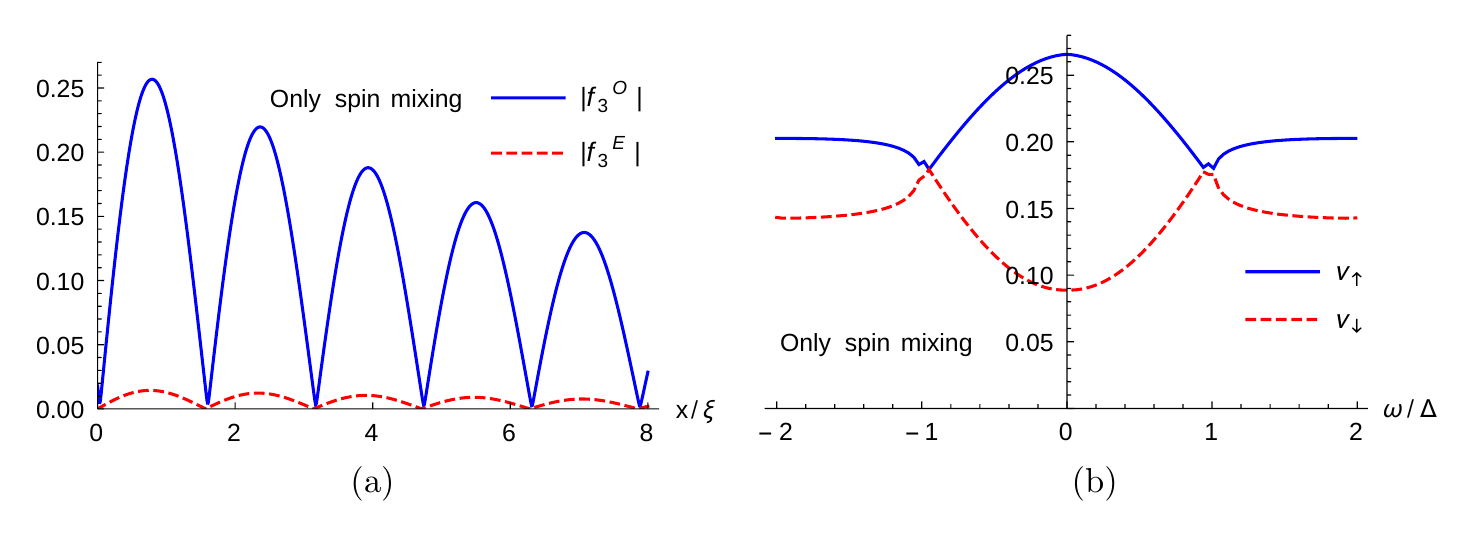}}
\caption{\small \sl{Absolute values of even and odd frequency mixed spin-triplet correlations induced in superconducting region for FS junction as function of position $x$ wherein only spin mixing occurs, (c) Frequency dependence of SPLDOS at $x=0$ for FS junction. Parameters are: $h/E_{F}=0.8$, $x'=0$, $\omega=0.1\Delta$ (for (a)), $E_{F}=10\Delta$.}}
\end{figure}
We see that when odd frequency mixed spin-triplet correlations dominate over even frequency mixed spin-triplet correlations, a peak is seen at $\omega=0$ for spin-up LDOS ($\nu_{\uparrow}$), while a dip is seen at $\omega=0$ for spin-down LDOS ($\nu_{\downarrow}$) at FS interface, while LMDOS is polarized in $z-$direction.
\subsubsection{Only spin-flip scattering occurs}
In our work in case of a Normal metal-Spin flipper-Superconductor junction we find that, $[G^{r}_{ee}]_{\uparrow\uparrow}=[G^{r}_{ee}]_{\downarrow\downarrow}$ and $[G^{r}_{ee}]_{\uparrow\downarrow}=[G^{r}_{ee}]_{\downarrow\uparrow}$ both in normal metal and superconducting region. Thus, from Eq.~\ref{lod}, we get,
\begin{equation}
\textbf{m} (x,\omega)=-\frac{2}{\pi}\text{Im}[[G^{r}_{ee}]_{\uparrow\downarrow}]\hat{x},\,\,\mbox{and}\,\,\,
\nu_{\sigma}=-\frac{1}{\pi}\lim_{\epsilon\rightarrow0}\text{Im}[[G^{r}_{ee}]_{\uparrow\uparrow}]+\frac{\sigma}{\pi} \lim_{\epsilon\to0}\text{Im}[[G^{r}_{ee}]_{\uparrow\downarrow}].
\label{lmsp}
\end{equation}
From Eq.~\eqref{lmsp} we see that only mixed ($\uparrow\downarrow$) spin component of the Green's function contributes to the LMDOS, while both equal ($\uparrow\uparrow$) and mixed ($\uparrow\downarrow$) spin components of the Green's function contribute to the SPLDOS. The mixed spin component of the Green's function, i.e., $[G^{r}_{ee}]_{\uparrow\downarrow}$ is finite only in presence of spin flip scattering. Thus, $[G^{r}_{ee}]_{\uparrow\downarrow}$ is responsible for the modification of the predicted SPLDOS at N-S interface. At N-S interface ($x=0$) for SPLDOS we find
\begin{equation}
\nu_{\sigma}=\frac{1}{2\pi}\text{Im}[\varrho_{1}]+\frac{\sigma}{2\pi}\sqrt{\text{Im}[\varrho_{2}]^2},
\label{spns}
\end{equation}
\begin{eqnarray}
\mbox{where,}\,\, \varrho_{1}=&&\frac{i\eta(2a_{81}k_{F}uv+b_{51}u^2(k_{F}-i\kappa)+b_{82}v^2(k_{F}+i\kappa)+(k_{F}-i(u^2-v^2)\kappa))}{(u^2-v^2)(k_{F}^2+\kappa^2)},\nonumber\\
\varrho_{2}=&&\frac{i\eta(b_{72}v^2(k_{F}+i\kappa)-2a_{62}k_{F}uv-b_{61}u^2(k_{F}-i\kappa))}{(u^2-v^2)(k_{F}^2+\kappa^2)}.\nonumber
\end{eqnarray}
In Eq.~\eqref{spns}, $\varrho_{2}$ is finite only in the presence of spin-flip scattering and contributes to the modification of SPLDOS at the NS interface. Further, from Eqs.~\eqref{tripletodd} and \eqref{spns} we see that both OTE-equal correlations in superconducting region and $\varrho_{2}$ depend on normal reflection amplitudes $b_{61}$, $b_{72}$ and Andreev reflection amplitude $a_{62}$. Thus large OTE-equal correlations in the superconducting region indicate large SPLDOS. Therefore the existence of OTE-equal correlations matches well with enhanced SPLDOS seen at NS interface.}

{Thus, in all three cases, LDOS is spin-polarized, or LMDOS is finite. Table I compares OTE-equal and OTE-mixed correlations with LMDOS and SPLDOS seen in these three cases. We find that both OTE-equal and OTE-mixed correlations are finite, and LMDOS is polarized in an arbitrary direction when both spin mixing and spin-flip scattering occur. When only spin mixing occurs, OTE-equal correlations vanish, but OTE-mixed correlations are finite with LMDOS polarized in $z$-direction. Finally, when only spin-flip scattering occurs, OTE-equal correlations are finite, but OTE-mixed correlations vanish with LMDOS polarized in $x$-direction.}
\begin{table}[ht]
{
\caption{{Comparison of OTE-equal, OTE-mixed, LMDOS and SPLDOS between three cases- (a) both spin mixing and spin flip scattering, (b) only spin mixing and, (c) only spin flip scattering}}
\begin{tabular}{|p{4.5cm}|p{1.2cm}|p{1.2cm}|p{2cm}|p{4.1cm}|p{4.1cm}|}
\hline
& OTE-equal & OTE-mixed & LMDOS & \multicolumn{2}{|c|}{SPLDOS}\\
\hline
& & & & $\omega=0$ & $\omega\simeq\pm\Delta$\\
\hline
Both spin mixing and spin flip scattering (F$_1$F$_2$S junction with misaligned magnetizations) & Finite & Finite & Polarized in any arbitrary direction& Peak in spin-up LDOS and dip in spin-down LDOS (Fig.~8(c)) & Dip in spin-up LDOS and peak in spin-down LDOS (Fig.~8(c))\\
\hline
Only spin mixing (NS junction with Rashba spin-orbit coupling or FS junction) & Zero & Finite & Polarized in $\hat{z}$-direction & Peak in spin-up LDOS ans dip in spin-down LDOS (Fig.~9(b)) & Dip in spin-up LDOS and peak in spin-down LDOS (Fig.~9(b)) \\
\hline
Only spin flip scattering (our case) & Finite & Zero & Polarized in $\hat{x}$-direction & Peak in \textbf{both} spin-up LDOS and spin-down LDOS for low values of spin flip scattering (Fig.~4(a)) & Dip in \textbf{both} spin-up LDOS and spin-down LDOS for low values of spin flip scattering (Fig.~4(a)) \\
\hline
\end{tabular}}
\end{table}
\section{Conclusion \& Perspective}
The setup as envisaged in Fig.~1 can be easily realized in a lab, as NS junctions have been in vogue for more than 40 years\cite{gusm}. Substituting a magnetic adatom or spin flipper at the NS interface shouldn't be difficult, especially with a $s$-wave superconductor like Lead or Aluminum; it should be perfectly possible.

To conclude, in this work, we have studied the emergence of odd frequency \textbf{equal} spin-triplet correlations at the interface of a Metal-Superconductor junction with a spin flipper. Using scattering Green's function approach, we have analytically calculated
even and odd frequency spin-singlet and equal spin-triplet correlations. Interestingly, we have found that in the presence of spin-flip scattering, mixed spin-triplet pairing vanishes, and only spin-singlet and equal spin-triplet pairings exist in our setup. In our normal metal-spin flipper-superconductor junction, we have observed that pairing correlations in the normal metal region show a nice oscillatory behavior at zero temperature. In contrast, at a finite temperature, they show an oscillatory decay. In superconducting region pairing correlations exhibit an oscillatory decay at both zero and finite temperatures. At low frequency and small values of spin-flip scattering, odd frequency equal spin-triplet correlations dominate over even frequency equal spin-triplet correlations in the superconducting region. It tallies with large values of the spin-polarized local density of states (SPLDOS) found for the same parameters. {We have also compared our obtained results for normal metal-spin flipper-superconductor junction with results from other hybrid junctions wherein either only spin mixing or both spin mixing and spin-flip scattering occurs. When only spin mixing occurs, odd frequency equal spin-triplet correlations vanish but odd frequency mixed spin-triplet correlations are finite. When both spin mixing and spin-flip scattering occur, both odd frequency equal spin-triplet correlations and odd frequency mixed spin-triplet correlations are finite. However, in the $N-SF-S$ junction, only spin-flip scattering is present, leading to finite odd frequency equal spin-triplet correlations with vanishing odd frequency mixed spin-triplet correlations.}

Odd/even frequency equal spin-triplet pairing hasn't yet been seen in a ballistic normal metal-$s$ wave superconductor junction; only odd/even frequency mixed spin-triplet correlation has been reported\cite{cayy,amb}. However, in this paper, we see evidence of odd/even frequency equal spin-triplet correlation in the presence of spin-flip scattering in a metal-superconductor junction. Shortly, we will study odd frequency equal spin-triplet pairing in a ferromagnetic Josephson junction in the presence of a spin-flipper\cite{anoodd}. We have already seen that a ferromagnetic Josephson junction embedded with a spin-flipper generates anomalous Josephson current, which is accompanied by quantized anomalous phase\cite{battery}. We will try to find possible relationships between odd frequency equal spin-triplet correlation and quantized anomalous phase.
\acknowledgments
This work was supported by the grants: 1. Josephson junctions with strained Dirac materials and their application in quantum information processing, SERB Grant No. CRG/20l9/006258, and 2. Nash equilibrium versus Pareto optimality in N-Player games, SERB MATRICS Grant No. MTR/2018/000070.
\appendix
\section{Analytical expressions for Green's functions in {N-SF-S junction}}
In this section we present analytical expressions for Green’s functions in both normal metal and superconducting
regions. These Green’s functions are used to calculate induced pairing correlations and SPLDOS in section IV.A of
our main text.
\subsection{Green's function in normal metal region}
Green's function in normal metal is obtained by plugging the wavefunctions from Eq.~\eqref{wave} for $x<0$ into Eq.~\eqref{RGF} with $b_{ij}$ and $a_{ij}$ found from Eqs.~\eqref{boun1}, \eqref{boun2}. For electron-electron and electron-hole components of Green's
function we get
\begin{equation}
\begin{split}
&[G^{r}_{ee}]_{\uparrow\uparrow}=-\frac{i\eta}{2k_{e}}[b_{11}e^{-ik_{e}(x+x')}+e^{ik_{e}|x-x'|}], \,\,\,\,
[G^{r}_{ee}]_{\downarrow\downarrow}=-\frac{i\eta}{2k_{e}}[b_{22}e^{-ik_{e}(x+x')}+e^{ik_{e}|x-x'|}],\\
&[G^{r}_{ee}]_{\uparrow\downarrow}=-\frac{i\eta}{2k_{e}}b_{21}e^{-ik_{e}(x+x')}, \,\,\,\,
[G^{r}_{ee}]_{\downarrow\uparrow}=-\frac{i\eta}{2k_{e}}b_{12}e^{-ik_{e}(x+x')},\\
&[G^{r}_{eh}]_{\uparrow\uparrow}=-\frac{i\eta}{2k_{e}}a_{31}e^{-i(k_{e}x-k_{h}x')}, \,\,\,\,
[G^{r}_{eh}]_{\downarrow\downarrow}=-\frac{i\eta}{2k_{e}}a_{42}e^{-i(k_{e}x-k_{h}x')},\\
&[G^{r}_{eh}]_{\uparrow\downarrow}=-\frac{i\eta}{2k_{e}}a_{41}e^{-i(k_{e}x-k_{h}x')}, \,\, \mbox{ and } \,\,
[G^{r}_{eh}]_{\downarrow\uparrow}=-\frac{i\eta}{2k_{e}}a_{32}e^{-i(k_{e}x-k_{h}x')}.
\end{split}
\end{equation}
We find that $b_{11}=b_{22}$, $b_{12}=b_{21}$, $a_{31}=-a_{42}=a_{11}$, $a_{41}=-a_{32}=a_{12}$. Therefore, we have
\begin{equation}
\label{nomamp}
\begin{split}
&[G^{r}_{ee}]_{\uparrow\uparrow}=[G^{r}_{ee}]_{\downarrow\downarrow}=-\frac{i\eta}{2k_{e}}[b_{11}e^{-ik_{e}(x+x')}+e^{ik_{e}|x-x'|}],\,\,
[G^{r}_{ee}]_{\uparrow\downarrow}=[G^{r}_{ee}]_{\downarrow\uparrow}=-\frac{i\eta}{2k_{e}}b_{12}e^{-ik_{e}(x+x')},\\
&[G^{r}_{eh}]_{\uparrow\uparrow}=-[G^{r}_{eh}]_{\downarrow\downarrow}=-\frac{i\eta}{2k_{e}}a_{11}e^{-i(k_{e}x-k_{h}x')},\,\, \mbox{ and }\,\,[G^{r}_{eh}]_{\uparrow\downarrow}=-[G^{r}_{eh}]_{\downarrow\uparrow}=-\frac{i\eta}{2k_{e}}a_{12}e^{-i(k_{e}x-k_{h}x')},\\\\
\mbox{ where, } b_{11}&=(-F^4J^4(u^2-v^2)^2-(y_{2}y_{3}(u^2-v^2)+J^2(1+m')^2(u^2-v^2)+yy_{3}u^2+yy_{2}v^2-iJ(1+m')(u^2-v^2)(y_{2}-y_{3}\\&+y-y_{1})+(y_{2}u^2+v^2(y_{3}-y)+u^2y)y_{1})(J^2m'^2(u^2-v^2)-y_{3}u^2y+y_{3}v^2y_{1}-u^2yy_{1}+v^2yy_{1}-iJm'(u^2-v^2)(y_{3}\\&+y+y_{1})+y_{2}(y_{3}(u^2-v^2)+iJm'(u^2-v^2)-v^2y+u^2y_{1}))-F^2J^2(y_{3}^2+2J^2m'(1+m')(u^2-v^2)^2+y_{2}^2\\&-y^2(u^2-v^2)^2+iJ(u^2-v^2)^2(y_{2}-y_{3}-y-2m'y-y_{1})+2y_{3}u^2(u^2-v^2)y_{1}+(u^2-v^2)^2y_{1}^2+2y_{2}v^2(4y_{3}u^2\\&+(u^2-v^2)y_{1})))/D_{S},\\
b_{12}&=2iFJy(y_{3}^2u^4+y_{2}^2u^2v^2+4y_{2}y_{3}u^2v^2+y_{3}^2u^2v^2+y_{2}^2v^4+F^2J^2(u^2-v^2)^2+J^2m'(1+m')(u^2-v^2)^2\\&+2 (u^2-v^2)(y_{3}u^2+y_{2}v^2)y_{1}+(u^2-v^2)^2y_{1}^2-iJ(u^2-v^2)(y_{3}u^2+v^2(y_{2}-y_{1})+u^2y_{1}))/D_{S},\\
a_{11}&=2FJ(y_{2}+y_{3})uv\sqrt{yy_{1}}(J(1+2m')(u^2-v^2)-i(y_{2}+y_{3}+(u^2-v^2)(y+y_{1})))/D_{S},\\
a_{12}&=2(y_{2}+y_{3})uv\sqrt{yy_{1}}(-J^2(F^2-(1+m')^2)(u^2-v^2)+yy_{3}u^2-iJ(1+m')(u^2-v^2)(y_{2}-y_{3}+y-y_{1})+y_{1}y_{3}v^2\\&+yy_{1}(u^2-v^2)+y_{2}(y_{3}(u^2-v^2)+v^2y+u^2y_{1}))/D_{S},\\
y=&\sqrt{1+\frac{\omega}{E_{F}}},\hspace{5pt}y_{1}=\sqrt{1-\frac{\omega}{E_{F}}}, \hspace{5pt}y_{2}=\sqrt{1+\frac{\sqrt{\omega^2-\Delta^2}}{E_{F}}}, \hspace{5pt}y_{3}=\sqrt{1-\frac{\sqrt{\omega^2-\Delta^2}}{E_{F}}},\\
D_{S}&=F^4J^4(u^2-v^2)^2+(y_{2}y_{3}(u^2-v^2)+J^2(1+m')^2(u^2-v^2)+yy_{3}u^2+yy_{2}v^2-iJ(1+m')(u^2-v^2)(y_{2}-y_{3}\\&+y-y_{1})+(y_{2}u^2+v^2(y_{3}-y)+u^2y)y_{1})(J^2m'^2(u^2-v^2)+yy_{3}u^2+y_{1}y_{3}v^2+yy_{1}(u^2-v^2)-iJm'(u^2-v^2)\\&(y_{3}-y+y_{1})+y_{2}(y_{3}(u^2-v^2)+iJm'(u^2-v^2)+v^2y+u^2y_{1}))+F^2J^2(y_{3}^2+2J^2m'(1+m')(u^2-v^2)^2\\&+y_{2}^2+2yy_{3}u^2v^2-2yy_{3}v^4+u^4y^2-2u^2v^2y^2+v^4y^2+iJ(u^2-v^2)^2(y_{2}-y_{3}+y-y_{1})+2y_{1}y_{3}u^2(u^2-v^2) \\&+(u^2-v^2)^2y_{1}^2+2y_{2}(4y_{3}u^2v^2+(u^2-v^2)(u^2y+v^2y_{1}))).
\end{split}
\end{equation}
Spin singlet and spin triplet pairing amplitudes are then calculated using Eq.~\eqref{pairingfunctions} in the main text, resulting in
\begin{equation}
f_{0}^{r}(x,x',\omega)=-\frac{i\eta}{2k_{e}}a_{12}e^{-i(k_{e}x-k_{h}x')},\,\,
f_{1}^{r}(x,x',\omega)=\frac{i\eta}{2k_{e}}a_{11}e^{-i(k_{e}x-k_{h}x')},\,\,
f_{2}^{r}(x,x',\omega)=0,\,\, \mbox{ and }\,\,
f_{3}^{r}(x,x',\omega)=0.
\end{equation}
In absence of spin flip scattering $b_{12}=a_{11}=0$, therefore from Eq.~\eqref{nomamp} we get
\begin{equation}
\begin{split}
&[G^{r}_{ee}]_{\uparrow\uparrow}=[G^{r}_{ee}]_{\downarrow\downarrow}=-\frac{i\eta}{2k_{e}}[b_{11}e^{-ik_{e}(x+x')}+e^{ik_{e}|x-x'|}],\,\,\,\,
[G^{r}_{ee}]_{\uparrow\downarrow}=[G^{r}_{ee}]_{\downarrow\uparrow}=0,\,\,\,\,
[G^{r}_{eh}]_{\uparrow\uparrow}=-[G^{r}_{eh}]_{\downarrow\downarrow}=0,\\
&[G^{r}_{eh}]_{\uparrow\downarrow}=-[G^{r}_{eh}]_{\downarrow\uparrow}=-\frac{i\eta}{2k_{e}}a_{12}e^{-i(k_{e}x-k_{h}x')}.
\end{split}
\end{equation}
wherein amplitudes $b_{11}$ and $a_{12}$ for no flip process can be obtained by putting $F=0$ in Eq.~\eqref{nomamp}. Spin singlet and spin triplet pairing amplitudes in absence of spin flip scattering are obtained from Eq.~\eqref{pairingfunctions} in main text, resulting in
\begin{equation}
f_{0}^{r}(x,x',\omega)=-\frac{i\eta}{2k_{e}}a_{12}e^{-i(k_{e}x-k_{h}x')},\,\,\,\,
f_{1}^{r}(x,x',\omega)=0,\,\,\,\,
f_{2}^{r}(x,x',\omega)=0,\,\, \mbox{ and }\,\,
f_{3}^{r}(x,x',\omega)=0,
\end{equation}
\subsection{Green's function in superconducting region}
In superconducting region we use same procedure as for normal metal region and finally get electron-electron and electron-hole components in presence of spin flip scattering as-
\begin{equation}
\label{supamp}
\begin{split}
[G^{r}_{ee}]_{\uparrow\uparrow}&=[G^{r}_{ee}]_{\downarrow\downarrow}=\frac{\eta}{2i(u^2-v^2)}\Bigg[\frac{e^{ik_{e}^{S}|x-x'|}u^2+b_{51}e^{ik_{e}^{S}(x+x')}u^2+a_{81}e^{i(k_{e}^{S}x'-k_{h}^{S}x)}uv}{k_{e}^{S}}\\&+\frac{a_{81}e^{i(k_{e}^{S}x-k_{h}^{S}x')}uv+b_{82}e^{-ik_{h}^{S}(x+x')}v^2+v^2e^{-ik_{h}^{S}|x-x'|}}{k_{h}^{S}}\Bigg],\\
[G^{r}_{ee}]_{\uparrow\downarrow}&=[G^{r}_{ee}]_{\downarrow\uparrow}=\frac{\eta}{2i(u^2-v^2)}\Bigg[\frac{b_{61}e^{ik_{e}^{S}(x+x')}u^2+a_{62}e^{i(k_{e}^{S}x'-k_{h}^{S}x)}uv}{k_{e}^{S}}-\frac{b_{72}e^{-ik_{h}^{S}(x+x')}v^2-a_{62}e^{i(k_{e}^{S}x-k_{h}^{S}x')}uv}{k_{h}^{S}}\Bigg],\\
[G^{r}_{eh}]_{\uparrow\uparrow}&=-[G^{r}_{eh}]_{\downarrow\downarrow}=-\frac{\eta}{2i(u^2-v^2)}\Bigg[\frac{b_{61}e^{ik_{e}^{S}(x+x')}uv+a_{62}e^{i(k_{e}^{S}x'-k_{h}^{S}x)}v^2}{k_{e}^{S}}+\frac{a_{62}e^{i(k_{e}^{S}x-k_{h}^{S}x')}u^2-b_{72}e^{-ik_{h}^{S}(x+x')}uv}{k_{h}^{S}}\Bigg],\\
[G^{r}_{eh}]_{\uparrow\downarrow}&=-[G^{r}_{eh}]_{\downarrow\uparrow}=\frac{\eta}{2i(u^2-v^2)}\Bigg[\frac{e^{ik_{e}^{S}|x-x'|}uv+b_{51}e^{ik_{e}^{S}(x+x')}uv+a_{81}e^{i(k_{e}^{S}x'-k_{h}^{S}x)}v^2}{k_{e}^{S}}\\&+\frac{a_{81}e^{i(k_{e}^{S}x-k_{h}^{S}x')}u^2+b_{82}e^{-ik_{h}^{S}(x+x')}uv+e^{-ik_{h}^{S}|x-x'|}uv}{k_{h}^{S}}\Bigg],\\\\
\mbox{ where, } b_{51}&=(-F^4J^4(u^2-v^2)^2-F^2J^2(y_{3}^2+2J^2m'(1+m')(u^2-v^2)^2-y_{2}^2+2yy_{3}u^2v^2-2yy_{3}v^4+y^2(u^2-v^2)^2\\&+2y_{1}y_{3}u^2(u^2-v^2)+(u^2-v^2)^2y_{1}^2-iJ(u^2-v^2)^2(y_{2}+y_{3}+2y_{2}m'-y+y_{1}))-(y_{2}y_{3}(u^2-v^2)\\&+J^2 (1+m')^2(u^2-v^2)+yy_{3}u^2+yy_{2}v^2-iJ(1+m')(u^2-v^2)(y_{2}-y_{3}+y-y_{1})+(y_{2}u^2+v^2(y_{3}-y)\\&+u^2y)y_{1})(J^2m'^2(u^2-v^2)+yy_{3}u^2+y_{1}y_{3}v^2+yy_{1}(u^2-v^2)-iJm'(u^2-v^2)(y_{3}-y+y_{1})-y_{2}(y_{3}(u^2-v^2)\\&+iJm'(u^2-v^2)+v^2y+u^2y_{1})))/D_{S},\\
b_{61}&=2iF'Jy_{2}(u^2-v^2)(y_{3}^2+F'^2J^2+J^2m'(m'-1)-2yy_{3}v^2+v^2y^2+2y_{1}y_{3}u^2+u^2y_{1}^2-iJ(y_{3}-v^2y+u^2y_{1}))/D_{S}^{\prime},\\
a_{62}&=-2iF'J\sqrt{y_{2}y_{3}}uv(u^2-v^2)(2F'^2J^2-2y_{2}y_{3}+2J^2m'(m'-1)+yy_{2}-yy_{3}+y^2+iJ(2y_{3}(m'-1)+2y_{2}m'\\&+y-y_{1})-y_{2}y_{1}+y_{1}y_{3}+y_{1}^2)/D_{S}^{\prime},\\
b_{72}&=-2iF'Jy_{3}(u^2-v^2)(y_{2}^2+F'^2J^2+J^2m'(m'-1)+2yy_{2}u^2+u^2y^2-2y_{1}y_{2}v^2+v^2y_{1}^2+iJ(y_{2}+u^2y-v^2y_{1}))/D_{S}^{\prime},\\
a_{81}&=-2\sqrt{y_{2}y_{3}}uv((y+y_{1})(y_{2}y_{3}(u^2-v^2)+J^2(1+m')^2(u^2-v^2)+yy_{3}u^2+yy_{2}v^2-iJ(1+m')(u^2-v^2)(y_{2}-y_{3}\\&+y-y_{1})+y_{1}y_{2}u^2+y_{1}y_{3}v^2+yy_{1}(u^2-v^2))+F^2J^2(2y_{2}+2y_{3}+(u^2-v^2)(y+y_{1})))/D_{S},\\
b_{82}&=(-F^4J^4(u^2-v^2)^2-(y_{2}y_{3}(u^2-v^2)+J^2(1+m1')^2(u^2-v^2)+yy_{3}u^2+yy_{2}v^2-iJ(1+m')(u^2-v^2)(y_{2}-y_{3}\\&+y-y_{1})+(y_{2}u^2+v^2(y_{3}-y)+u^2y)y_{1})(J^2m'^2(u^2-v^2)-yy_{3}u^2+iJm'(u^2-v^2)(y_{3}+y-y_{1})-y_{1}y_{3}v^2\\&+yy_{1}(u^2-v^2)+y_{2}(iJm'(u^2-v^2)+y_{3}(-u^2+v^2)+v^2y+u^2y_{1}))-F^2J^2(-y_{3}^2+2J^2m'(1+m')(u^2-v^2)^2\\&+y_{2}^2+y^2(u^2-v^2)^2+iJ (u^2-v^2)^2(y_{2}+y_{3}+2y_{3}m'+y-y_{1})+(u^2-v^2)^2y_{1}^2+2y_{2}(u^2-v^2)(u^2y+v^2y_{1})))/D_{S},\\
D_{S}^{\prime}&=F'^4J^4(u^2-v^2)^2+(y_{2}y_{3}(u^2-v^2)+J^2m'^2(u^2-v^2)+yy_{3}u^2+yy_{2}v^2-iJm'(u^2-v^2)(y_{2}-y_{3}\\&+y-y_{1})+(y_{2}u^2+v^2(y_{3}-y)+u^2y)y_{1})(J^2(m'-1)^2(u^2-v^2)+yy_{3}u^2+y_{1}y_{3}v^2+yy_{1}(u^2-v^2)-iJ(m'-1)\\&(u^2-v^2)(y_{3}-y+y_{1})+y_{2}(y_{3}(u^2-v^2)+iJ(m'-1)(u^2-v^2)+v^2y+u^2y_{1}))+F'^2J^2(y_{3}^2+2J^2m'(m'-1)\\&(u^2-v^2)^2+y_{2}^2+2yy_{3}u^2v^2-2yy_{3}v^4+u^4y^2-2u^2v^2y^2+v^4y^2+iJ(u^2-v^2)^2(y_{2}-y_{3}+y-y_{1})\\&+2y_{1}y_{3}u^2(u^2-v^2)+(u^2-v^2)^2y_{1}^2+2y_{2}(4y_{3}u^2v^2+(u^2-v^2)(u^2y+v^2y_{1}))).
\end{split}
\end{equation}
From the anomalous electron-hole component of retarded Green's function we get spin singlet and spin triplet pairing amplitudes using Eq.~\eqref{pairingfunctions} as-
\begin{eqnarray}
\begin{split}
f_{0}^{r}&=\frac{\eta uv}{2i(u^2-v^2)}\Bigg\{e^{-\kappa|x-x'|}\Bigg[\frac{e^{ik_{F}|x-x'|}}{k_{e}^{S}}+\frac{e^{-ik_{F}|x-x'|}}{k_{h}^{S}}\Bigg]+e^{-\kappa(x+x')}\Bigg[\frac{b_{51}e^{ik_{F}(x+x')}}{k_{e}^{S}}+\frac{b_{82}e^{-ik_{F}(x+x')}}{k_{h}^{S}}\Bigg]\\&+a_{81}e^{-\kappa(x+x')}\Bigg[\frac{ue^{ik_{F}(x-x')}}{vk_{h}^{S}}+\frac{ve^{-ik_{F}(x-x')}}{uk_{e}^{S}}\Bigg]\Bigg\},\\
f_{1}^{r}&=\frac{\eta uv}{2i(u^2-v^2)}\Bigg\{e^{-\kappa(x+x')}\Bigg[\frac{b_{61}e^{ik_{F}(x+x')}}{k_{e}^{S}}-\frac{b_{72}e^{-ik_{F}(x+x')}}{k_{h}^{S}}\Bigg]+a_{62}e^{-\kappa(x+x')}\Bigg[\frac{ue^{ik_{F}(x-x')}}{vk_{h}^{S}}+\frac{ve^{-ik_{F}(x-x')}}{uk_{e}^{S}}\Bigg]\Bigg\},\\
f_{2}^{r}&=0,\,\,\,\,
f_{3}^{r}=0.
\end{split}
\end{eqnarray}
In absence of spin flip scattering $b_{61}=b_{72}=a_{62}=0$, therefore from Eq.~\eqref{supamp} we obtain
\begin{equation}
\begin{split}
[G^{r}_{ee}]_{\uparrow\uparrow}&=[G^{r}_{ee}]_{\downarrow\downarrow}=\frac{\eta}{2i(u^2-v^2)}\Bigg[\frac{e^{ik_{e}^{S}|x-x'|}u^2+b_{51}e^{ik_{e}^{S}(x+x')}u^2+a_{81}e^{i(k_{e}^{S}x'-k_{h}^{S}x)}uv}{k_{e}^{S}}\\&+\frac{a_{81}e^{i(k_{e}^{S}x-k_{h}^{S}x')}uv+b_{82}e^{-ik_{h}^{S}(x+x')}v^2+v^2e^{-ik_{h}^{S}|x-x'|}}{k_{h}^{S}}\Bigg],\\
[G^{r}_{ee}]_{\uparrow\downarrow}&=[G^{r}_{ee}]_{\downarrow\uparrow}=0,\,\,\,\,
[G^{r}_{eh}]_{\uparrow\uparrow}=-[G^{r}_{eh}]_{\downarrow\downarrow}=0,\\
[G^{r}_{eh}]_{\uparrow\downarrow}&=-[G^{r}_{eh}]_{\downarrow\uparrow}=\frac{\eta}{2i(u^2-v^2)}\Bigg[\frac{e^{ik_{e}^{S}|x-x'|}uv+b_{51}e^{ik_{e}^{S}(x+x')}uv+a_{81}e^{i(k_{e}^{S}x'-k_{h}^{S}x)}v^2}{k_{e}^{S}}\\&+\frac{a_{81}e^{i(k_{e}^{S}x-k_{h}^{S}x')}u^2+b_{82}e^{-ik_{h}^{S}(x+x')}uv+e^{-ik_{h}^{S}|x-x'|}uv}{k_{h}^{S}}\Bigg],\\
\end{split}
\end{equation}
expressions for $b_{51}$, $b_{82}$ and $a_{81}$ in absence of spin flip scattering can be found by putting $F=0$ in Eq.~\eqref{supamp}. Finally, the spin singlet and spin triplet pairing amplitudes for no flip process are given as from Eq.~\eqref{pairingfunctions} in main text
\begin{equation}
\begin{split}
&f_{0}^{r}=\frac{\eta uv}{2i(u^2-v^2)}\Bigg\{e^{-\kappa|x-x'|}\Bigg[\frac{e^{ik_{F}|x-x'|}}{k_{e}^{S}}+\frac{e^{-ik_{F}|x-x'|}}{k_{h}^{S}}\Bigg]+e^{-\kappa(x+x')}\Bigg[\frac{b_{51}e^{ik_{F}(x+x')}}{k_{e}^{S}}+\frac{b_{82}e^{-ik_{F}(x+x')}}{k_{h}^{S}}\Bigg]\\&+a_{81}e^{-\kappa(x+x')}\Bigg[\frac{ue^{ik_{F}(x-x')}}{vk_{h}^{S}}+\frac{ve^{-ik_{F}(x-x')}}{uk_{e}^{S}}\Bigg]\Bigg\},\\
&\mbox{while, }f_{1}^{r}=0,\,\,\,\, f_{2}^{r}=0,\,\,\,\, f_{3}^{r}=0.
\end{split}
\end{equation}
\section{N-SF-S junction (Finite temperature)}
In section IV.A we have discussed spin singlet and triplet correlations at zero temperature. In this section we will study effect of finite temperature on spin singlet and triplet correlations. To calculate correlations at finite temperature we use Matsubara representation, replacing $\omega$ with $i\omega_{n}$ in anomalous electron-hole propagator (see Eqs.~\eqref{green}, \eqref{greenFT}). In electron-hole propagator at finite temperature\eqref{greenFT}, summation is taken over positive frequencies only because all pairing correlations become odd functions of frequency. From electron-hole propagator at finite temperature (Eq.~\eqref{greenFT}), we can compute even/odd frequency spin singlet and spin triplet correlations, see Eqs~\eqref{singleteven}, \eqref{singletodd}, \eqref{tripleteven}, \eqref{tripletodd}.
\subsection{Odd versus even frequency spin-singlet correlations}
At zero temperature, both even and odd frequency spin-singlet correlations in the N region exhibit an oscillatory behavior and survive infinitely far away. However, at finite temperature ($T$), ESE and OSO correlations oscillate as well as decay in N region since decay length in N region is $\xi_{N}\sim\frac{1}{T}$\cite{cov,sdo}.

Even and odd frequency spin singlet correlations at finite temperature are given as-
\begin{equation}
\label{singletevenoddFT}
f_{0}^{E}(x,x',T)=\sum_{\omega_{n}>0}f_{0}^{E}(x,x',\omega \to i\omega_{n}),\, \mbox{ and }\,
f_{0}^{O}(x,x',T)=\sum_{\omega_{n}>0}f_{0}^{O}(x,x',\omega \to i\omega_{n})
\end{equation}
where $f_{0}^{E}(x,x',\omega)$ and $f_{0}^{O}(x,x',\omega)$ are given in Eqs.~\eqref{singleteven}, \eqref{singletodd}.
In Fig.~10 we plot spin singlet correlation induced in N($x<0$) and S($x>0$) regions as a function of position $x$ at finite temperature for both no flip (Fig.~10(a)) and spin flip (Figs.~10(b,c)) processes. Even and odd frequency spin singlet pairings are finite and show a nice oscillatory decay as function of position $x$ in the normal region.
\begin{figure}[h]
\centering{\includegraphics[width=.99\textwidth]{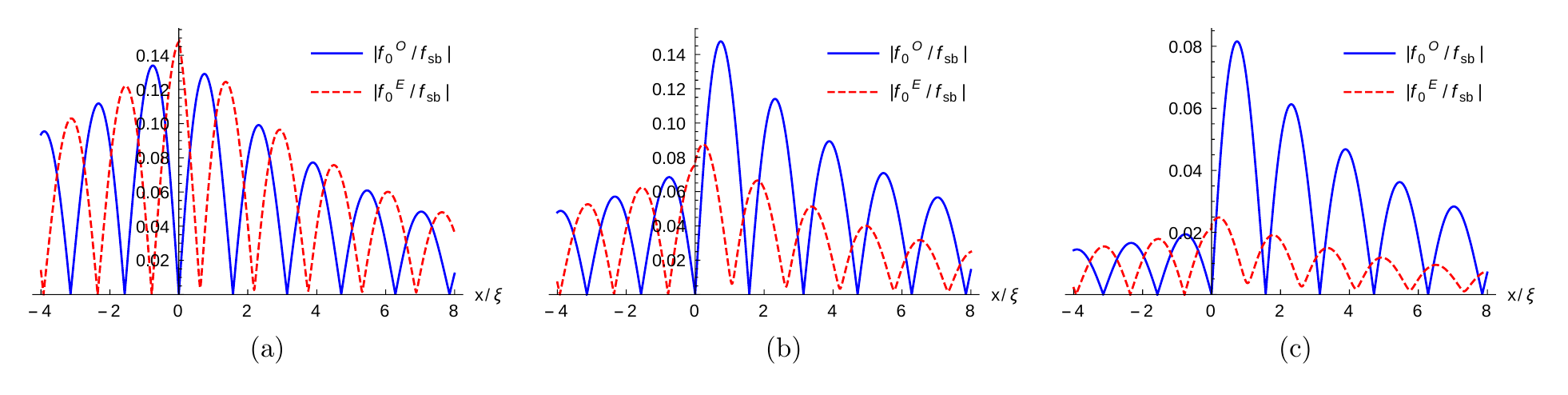}}
\caption{\small \sl The absolute values of the even and odd frequency spin-singlet correlation induced in the N region ($x<0$) and S region ($x>0$) as a function of the position $x$ for (a) no flip process and (b) spin flip process. Parameters are: $S=1/2$ (for (a) and (b)), $S=5/2$ (for (c)), $F=F'=0$ (for (a)), $F=F'=1$ (for (b)), $F=F'=3$ (for (c)), $J=1$, $x'=0$, $T/T_{c}=0.01$, $E_{F}=10\Delta$.}
\end{figure}
The reason for this kind of behavior can be understood by substituting $\omega$ with $i\omega_{n}$ in Eqs.~\eqref{singleteven}, \eqref{singletodd} where ESE correlation is proportional to $e^{k^{M'}(x+x')}\cos[k_{F}(x-x')]$ and OSO correlation is proportional to $e^{k^{M'}(x+x')}\sin[k_{F}(x-x')]$ in N region ($x<0$), with $k^{M'}=\omega_{n} k_{F}/(2E_{F})$. This is in contrast to what we observe at zero temperature where ESE and OSO correlations exhibit a nice oscillation instead of oscillatory decay at zero temperature. In S region, we see a nice oscillatory decay similar to zero temperature, only the magnitudes of pairing correlations may change but qualitatively there is no change when $\omega\rightarrow i\omega_{n}$ since the factor $\kappa$ ($=\sqrt{(\Delta^2-\omega^2)}[k_{F}/(2E_{F})]$) occurring in the superconducting wavefunctions is the function of $\omega^2$.

In our figures we normalize the pairing amplitudes to the value of spin singlet pairing amplitude in the bulk superconductors\cite{ltr},
\begin{equation}
f_{sb}=2\sum_{\omega_{n}}\frac{\Delta}{\sqrt{\omega_{n}^2+\Delta^2}}.
\end{equation}
The temperature dependence of the bulk pair potential $\Delta$ is given as $\Delta(T)=\Delta(0)\tanh(1.74\sqrt{T_{c}/T-1})$, where $T_{c}$ is the critical temperature\cite{aun}.
\subsection{Odd versus even frequency equal spin triplet correlations}
Finite temperature, even and odd frequency equal spin triplet correlations ($f_{\uparrow\uparrow}^{E}(x,x',T), f_{\uparrow\uparrow}^{O}(x,x',T)$) are derived by substituting $i\omega_{n}$ for $\omega$ in Eqs.~\eqref{tripleteven}, \eqref{tripletodd}. Thus
\begin{eqnarray}
\label{tripletevenFT}
&&f_{\uparrow\uparrow}^{E}(x,x',T)=-f_{\downarrow\downarrow}^{E}(x,x',T)=\sum_{\omega_{n}>0}f_{\uparrow\uparrow}^{E}(x,x',\omega \to i\omega_{n})=-\sum_{\omega_{n}>0}f_{\downarrow\downarrow}^{E}(x,x',\omega \to i\omega_{n}),\\
\label{tripletoddFT}
&&f_{\uparrow\uparrow}^{O}(x,x',T)=-f_{\downarrow\downarrow}^{O}(x,x',T)=\sum_{\omega_{n}>0}f_{\uparrow\uparrow}^{O}(x,x',\omega \to i\omega_{n})=-\sum_{\omega_{n}>0}f_{\downarrow\downarrow}^{O}(x,x',\omega \to i\omega_{n}),
\end{eqnarray}
in presence of spin flip scattering. Similar to zero temperature case, in absence of spin flip scattering equal spin triplet correlations (Eq.~\eqref{tripletevenFT}, \eqref{tripletoddFT}) vanish.

In Fig.~11 we plot ETO-equal ($f_{\uparrow\uparrow}^{E}$, $f_{\downarrow\downarrow}^{E}$) and OTE-equal ($f_{\uparrow\uparrow}^{O}$, $f_{\downarrow\downarrow}^{O}$) correlations as a function of position $x$ for small ($F=F'=1$, Fig.~11(a)) and large ($F=F'=3$, Fig.~11(b)) values of spin flip scattering. We see that equal spin triplet correlations are finite and exhibit an oscillatory decay in N region. This is in contrast to what we see for equal spin triplet correlations at zero temperature, see Figs.~3(a), 3(b).
\begin{figure}[h]
\centering{\includegraphics[width=.99\textwidth]{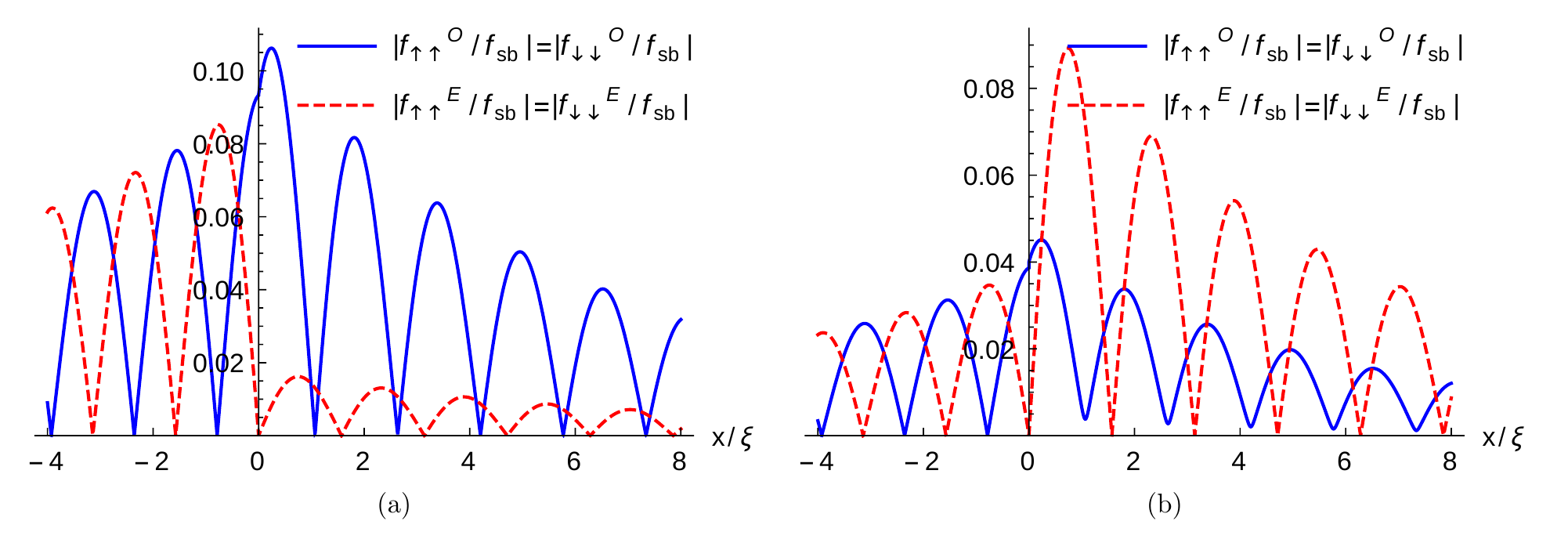}}
\caption{\small \sl Absolute values of even and odd frequency equal spin-triplet correlation induced in N region ($x<0$) and S region ($x>0$) as a function of position $x$ for spin flip process. Parameters are: $S=1/2$ (for (a)), $S=5/2$ (for (b)), $F=F'=1$ (for (a)), $F=F'=3$ (for (b)), $J=1$, $x'=0$, $T/T_{c}=0.01$, $E_{F}=10\Delta$.}
\end{figure}
The reason for this behavior can be understood by substituting $\omega$ with $i\omega_{n}$ in Eqs.~\eqref{tripleteven}, \eqref{tripletodd} where ETO-equal correlation is proportional to $e^{k^{M'}(x+x')}\sin[k_{F}(x-x')]$ and OTE-equal correlation is proportional to $e^{k^{M'}(x+x')}\cos[k_{F}(x-x')]$ in the N region ($x<0$). In S region we see the similar behavior for both ETO-equal ($f_{\uparrow\uparrow}^{E}$, $f_{\downarrow\downarrow}^{E}$) and OTE-equal ($f_{\uparrow\uparrow}^{O}$, $f_{\downarrow\downarrow}^{O}$) correlations as seen at zero temperature. Finally, we note that both even as well as odd frequency mixed spin triplet correlations vanish regardless of spin flip scattering, i.e., $f_{3}^{E}(x,x',T)=f_{3}^{O}(x,x',T)=0$. This is an unique result of our work since most papers report odd frequency mixed spin triplet correlations with vanishing odd frequency equal spin triplet correlations\cite{amb,cayy}.
\section{Analytical expressions for Green's functions in F${}_{1}$-F${}_{2}$-S junction}
In Appendixes A and B we have provided analytical expressions for Green’s functions and shown the effect of finite
temperature on spin singlet and triplet correlations in case of N-SF-S junction respectively. In this Appendix we give
analytical expressions for Green’s functions in case of F${}_{1}$-F${}_{2}$-S junction. These Green’s functions are used to compute
induced even/odd frequency spin singlet, spin triplet correlations and, SPLDOS in section V of our main text. In
section V, we have explained how and why our obtained results differ from those obtained previously. In this regard,
we have calculated induced pairing correlations as well as SPLDOS in case of F${}_{1}$-F${}_{2}$-S junction wherein both spin flip
scattering and spin mixing occur. Herein below we show analytical expressions for Green’s functions and how both
equal and mixed spin triplet correlations are finite in left ferromagnetic region.

{Green's function in left ferromagnetic region is obtained by plugging the wavefunctions from Eq.~\eqref{wavfero} for $x<-a$ into Eq.~\eqref{RGF} with $b_{ij}^{\prime}$ and $a_{ij}^{\prime}$ found from Eqs.~\eqref{BCF1}-\eqref{BCF2}. For electron-electron and electron-hole components of Green's function we get,
\begin{equation}
\begin{split}
&[G^{r}_{ee}]_{\uparrow\uparrow}=-\frac{i\eta}{2q_{\uparrow}^{+}}[b'_{11}e^{-iq_{\uparrow}^{+}(x+x')}+e^{iq_{\uparrow}^{+}|x-x'|}], \,\,\,\,
[G^{r}_{ee}]_{\downarrow\downarrow}=-\frac{i\eta}{2q_{\downarrow}^{+}}[b'_{22}e^{-iq_{\downarrow}^{+}(x+x')}+e^{iq_{\downarrow}^{+}|x-x'|}],\\
&[G^{r}_{ee}]_{\uparrow\downarrow}=-\frac{i\eta}{2q_{\downarrow}^{+}}b'_{21}e^{-i(q_{\uparrow}^{+}x+q_{\downarrow}^{+}x')}, \,\,\,\,
[G^{r}_{ee}]_{\downarrow\uparrow}=-\frac{i\eta}{2q_{\uparrow}^{+}}b'_{12}e^{-i(q_{\downarrow}^{+}x+q_{\uparrow}^{+}x')},\\
&[G^{r}_{eh}]_{\uparrow\uparrow}=-\frac{i\eta}{2q_{\uparrow}^{-}}a'_{31}e^{-i(q_{\uparrow}^{+}x-q_{\uparrow}^{-}x')}, \,\,\,\,
[G^{r}_{eh}]_{\downarrow\downarrow}=-\frac{i\eta}{2q_{\downarrow}^{-}}a'_{42}e^{-i(q_{\downarrow}^{+}x-q_{\downarrow}^{-}x')},\\
&[G^{r}_{eh}]_{\uparrow\downarrow}=-\frac{i\eta}{2q_{\downarrow}^{-}}a'_{41}e^{-i(q_{\uparrow}^{+}x-q_{\downarrow}^{-}x')}, \,\, \mbox{ and } \,\,
[G^{r}_{eh}]_{\downarrow\uparrow}=-\frac{i\eta}{2q_{\uparrow}^{-}}a'_{32}e^{-i(q_{\downarrow}^{+}x-q_{\uparrow}^{-}x')}.
\end{split}
\label{grefer}
\end{equation}
Substituting Eq.~\eqref{grefer} in Eq.~\eqref{pairingfunctions} we get,
\begin{equation}
\begin{split}
&f_{0}^{r}(x,x',\omega)=-\frac{i\eta}{4}\Bigg(\frac{a'_{41}e^{-i(q_{\uparrow}^{+}x-q_{\downarrow}^{-}x')}}{q_{\downarrow}^{-}}-\frac{a'_{32}e^{-i(q_{\downarrow}^{+}x-q_{\uparrow}^{-}x')}}{q_{\uparrow}^{-}}\Bigg),\,\,
f_{1}^{r}(x,x',\omega)=-\frac{i\eta}{4}\Bigg(\frac{a'_{42}e^{-i(q_{\downarrow}^{+}x-q_{\downarrow}^{-}x')}}{q_{\downarrow}^{-}}-\frac{a'_{31}e^{-i(q_{\uparrow}^{+}x-q_{\uparrow}^{-}x')}}{q_{\uparrow}^{-}}\Bigg),\\
&f_{2}^{r}(x,x',\omega)=-\frac{\eta}{4}\Bigg(\frac{a'_{31}e^{-i(q_{\uparrow}^{+}x-q_{\uparrow}^{-}x')}}{q_{\uparrow}^{-}}+\frac{a'_{42}e^{-i(q_{\downarrow}^{+}x-q_{\downarrow}^{-}x')}}{q_{\downarrow}^{-}}\Bigg), \mbox{ and }
f_{3}^{r}(x,x',\omega)=-\frac{i\eta}{4}\Bigg(\frac{a'_{41}e^{-i(q_{\uparrow}^{+}x-q_{\downarrow}^{-}x')}}{q_{\downarrow}^{-}}+\frac{a'_{32}e^{-i(q_{\downarrow}^{+}x-q_{\uparrow}^{-}x')}}{q_{\uparrow}^{-}}\Bigg).
\end{split}
\end{equation}
Thus, equal spin triplet correlations $f_{\uparrow\uparrow}$ and $f_{\downarrow\downarrow}$ are finite, $f_{\uparrow\uparrow}=if_{2}^{r}-f_{1}^{r}=-\frac{i\eta}{2q_{\uparrow}^{-}}a'_{31}e^{-i(q_{\uparrow}^{+}x-q_{\uparrow}^{-}x')}$, and $f_{\downarrow\downarrow}=if_{2}^{r}+f_{1}^{r}=-\frac{i\eta}{2q_{\downarrow}^{-}}a'_{42}e^{-i(q_{\downarrow}^{+}x-q_{\downarrow}^{-}x')}$ and mixed spin triplet correlations $f_{3}^{r}$ are also finite.}

\end{document}